\documentclass[pre,longbibliography,twocolumn]{revtex4-1}
\usepackage[utf8]{inputenc}
\usepackage{natbib}
\usepackage{amssymb,amsmath}
\usepackage{epsfig}
\usepackage{bm}
\usepackage{soul}
\usepackage{xcolor}

\usepackage{graphicx}

\newcommand\beq{\begin{equation}}
\newcommand\eeq{\end{equation}}
\newcommand\beqa{\begin{eqnarray}}
\newcommand\eeqa{\end{eqnarray}}
\newcommand{\dd}{\text{d}}

\newcommand{\al}{\alpha}

\newcommand*\Bell{\ensuremath{\boldsymbol\ell}}

\newcommand{\gammaSt}{\gamma_{\scalebox{1.1}{$\scriptscriptstyle \text{St}$}}}
\newcommand{\elee}{\ell_\text{e}}

\begin{document}

\title{Diffusion of intruders in granular suspensions: Enskog theory and random walk interpretation}

\author{Rub\'en G\'omez Gonz\'alez\footnote[1]{Electronic address: ruben@unex.es}}
\affiliation{Departamento de F\'{\i}sica,
Universidad de Extremadura, E-06006 Badajoz, Spain}
\author{Enrique Abad\footnote[2]{Electronic address: eabad@unex.es}}
\affiliation{Departamento de F\'{\i}sica Aplicada and Instituto de Computaci\'on Cient\'{\i}fica Avanzada (ICCAEx), Universidad de Extremadura, 06800 M\'erida, Spain}
\author{Santos Bravo Yuste\footnote[3]{Electronic address: santos@unex.es} and Vicente Garz\'o\footnote[4]{Electronic address: vicenteg@unex.es;
URL: http://www.unex.es/eweb/fisteor/vicente/}}
\affiliation{Departamento de F\'{\i}sica and Instituto de Computaci\'on Cient\'{\i}fica Avanzada (ICCAEx), Universidad de Extremadura, E-06006 Badajoz, Spain}

\begin{abstract}
The Enskog kinetic theory is applied to compute the mean square displacement of impurities or intruders (modeled as smooth inelastic hard spheres)
immersed in a granular gas of smooth inelastic hard spheres (grains). Both species (intruders and grains) are surrounded by an interstitial molecular gas (background) that plays the role of a thermal bath. The influence of the latter on the motion of intruders and grains is modeled via a standard viscous drag force supplemented by a stochastic Langevin-like force proportional to the background temperature. 
We solve the corresponding Enskog-Lorentz kinetic
equation by means of the Chapman-Enskog expansion truncated to first order in the gradient of the intruder number density.  The integral equation for the diffusion coefficient is solved by considering the first two Sonine approximations.
To test these results,
we also compute the diffusion coefficient  from the numerical solution of the inelastic Enskog equation by means of the direct simulation Monte Carlo method.  We find that the first Sonine approximation generally agrees
well with the simulation results, although significant discrepancies arise when the intruders become lighter
than the grains. Such discrepancies are largely mitigated by the use of the second-Sonine approximation, in excellent agreement with computer simulations even for moderately strong inelasticities and/or dissimilar mass and diameter ratios. We invoke a random walk
picture of the intruders' motion to shed light on the physics underlying the intricate dependence
of the diffusion coefficient on the main system parameters.
This approach, recently employed to study the case of an intruder immersed in a granular gas, also
proves useful in the present case of a granular suspension.
Finally, we discuss the applicability of our model to real systems
in the self-diffusion case. We conclude that collisional effects may strongly impact the diffusion coefficient of the grains.
\end{abstract}

\draft
\date{\today}
\maketitle

\section{Introduction}
\label{sec1}

Granular systems are constituted by macroscopic particles (or ``grains'') that collide inelastically with one another, implying that their total kinetic energy decreases in time. Such freely cooling granular systems exhibit interesting nonequilibrium transport properties at macroscopic scales, such as slowed-down diffusion \cite{BDPB02,MJChB14,BChChM15}.
However, because of the lack of homogeneity induced by gravity, boundary effects and the onset of clustering instabilities, it is quite difficult to confirm by experiments the theoretical predictions for the dynamic properties of the
so-called homogeneous cooling state (HCS). We note nonetheless that interesting observations concerning the fulfillment
of Haff's law in the HCS of a granular gas under microgravity conditions have been published \cite{TMHS09,HTWS18,YSS20}.

In order to observe sustained diffusive motion on much longer time scales, an external energy input is required to maintain
the system under rapid flow conditions. A non-equilibrium steady state is reached
when the external energy supplied exactly counterbalances the kinetic energy loss by grain-grain collisions.
In real experiments, a variety of mechanisms can be used to inject energy into the system at hand, e.g. mechanical-boundary shaking \cite{YHCMW02,HYCMW04}, bulk driving (as in air-fluidized beds \cite{SGS05,AD06}), or magnetic forces \cite{SHKZP13,HTWS18}.
In most cases, the formation of large spatial gradients in the bulk domain becomes unavoidable; consequently, a rigorous theoretical
description must go beyond the Navier-Stokes framework, and one is then confronted with great difficulties. In computer simulations,
this obstacle can be circumvented by the introduction of external forces (or \emph{thermostats}) that heat the system and compensate for the
energy dissipated by collisions. Unfortunately, in most cases it is not clear how to realize each specific
type of thermostat in experiments. That said, some thermostats appear to be less artifactual and more physically
transparent than others. In particular, a more realistic example of thermostated granular systems consists of a set of solid particles immersed
in an interstitial fluid of molecular particles. This provides a suitable starting point to mimic the behavior of real suspensions.

Needless to say, understanding the flow of solid particles in one or more fluid phases is a very intricate problem not only from a fundamental
point of view, but also from a practical perspective. A prominent example among the different types of gas-solid flows
are the so-called particle-laden suspensions, consisting of a (typically dilute) collection of small grains immersed in a carrier fluid \cite{S20}. For a proper parameter choice, the dynamics of grains are essentially dominated by collisions between them,
and the tools of kinetic theory conveniently adapted to account for the inelastic character of the collisions can be employed
successfully to describe this type of granular flows \cite{S20,RN08}.

Because of the aforementioned complexity embodied in the description of two or more phases, a coarse-grained approach is usually adopted. In this context, the effect of the interstitial fluid (background) on grains is usually incorporated
in an effective way through a fluid-solid interaction force $\mathbf{F}_\text{fluid}$ \cite{K90,G94,J00,KH01}. Some of the continuum approaches of gas-solid flows \cite{SJ89,CBC23} have been based on the addition of an empirical drag law to the solid momentum balance without any gas-phase modifications in the transport coefficients of the solid phase. On the other hand, in the past few years the impact of the interstitial gas on the transport properties of grains has been incorporated with increasing rigor. In this context, the effect of the interstitial gas phase on grains in the continuum derivation is usually accounted for from the very beginning via  the (inelastic) Enskog equation for the solid phase (see representative reviews in Refs.\ \cite{KH01,FH17,CW24}). In particular, numerous groups \cite{LMJ91,TK95,SMTK96,WZLH09,PS12,H13,WGZS14,SA17,ASG19,SA20} have implemented this approach by incorporating a viscous drag force proportional to the instantaneous particle velocity. This drag force aims to account for the friction exerted on the grains by the interstitial fluid. However, some works \cite{TGHFS10} have revealed that the drag force term fails to capture the particle acceleration-velocity correlation observed in direct numerical simulations \cite{TS14}. Based on the results of the latter, we have chosen to remedy the aforementioned shortcoming by
including a stochastic Langevin-like term in the effective force $\mathbf{F}_\text{fluid}$. This new contribution mimics the additional effects of neighboring particles by means of a noise term involving the stochastic increment of a Wiener process \cite{GTSH12}. Specifically, this stochastic term (which \emph{randomly} kicks the particles between collisions) accounts for the energy transfer from the background particles to the granular gas. Thus, the gas-phase effect is described by
the addition of a Fokker-Planck term in the final Enskog equation.

It is worth emphasizing that the above suspension model \cite{GTSH12} is not purely heuristic, as it can actually be derived in a rigorous way from a more detailed level of description, i.e., by explicit consideration of the (elastic) collisions between grains and the molecular gas particles. To this end, such collisions are modeled with the help of the Boltzmann--Lorentz collision operator \cite{RL77}. While this sort of suspension model applies for arbitrary mass ratios of granular and molecular gases, a simplification occurs when the grains are much heavier than the particles of the molecular gas. In this limiting case, the Boltzmann--Lorentz operator reduces to the Fokker--Planck operator, and one indeed finds full agreement between the results for transport properties derived from the collisional model \cite{GG22} and those obtained from a coarse-grained approach \cite{GGG19a}.

Although the use of effective forces for modeling gas-solid flows is quite common in the granular literature, one should not lose
sight of the assumptions involved in this sort of derivation. First, since the granular particle particles are sufficiently dilute, we
assume that the state of the interstitial gas is not affected by the presence of the grains, implying that the former can be treated as a
\emph{thermostat} at a constant temperature $T_\text{b}$. Second,
we assume that the effect of the background gas on grain-grain collisions can be neglected, so that  for moderately dense gases the
state of grains is mainly determined by
collisions between themselves. Consequently, for moderately
dense gases, the Enskog collision operator takes the same form as that of a \emph{dry} granular gas \cite{G19}. This implies that the collision dynamics lacks any parameter associated with the surrounding gas. As discussed in several previous papers \cite{KH01,K90,TK95,SMTK96,WKL03}, the above assumption requires the mean-free time between collisions to be substantially less than the time required by gas-solid forces to significantly impact the dynamics of solid particles. This requirement is clearly fulfilled in scenarios where the motion of grains is minimally influenced by the gas phase; however, it no longer holds in other situations (e.g., when solid particles are immersed in a liquid)  since the surrounding fluid strongly affects the collision process. A final simplifying assumption will consist in restricting our analysis to the regime of low Reynolds numbers (Stokes flow); in this regime, the inertia of the fluid is negligible compared to its viscous forces.

In the sequel, we will consider a dilute granular suspension (a system constituted by the background gas and the grains) subject to the above assumptions. For the grains we consider in this paper (i.e., for smooth inelastic  hard spheres), the inelasticity of the binary collisions is quantified by the  so-called coefficient of normal restitution $\alpha$. For two colliding spheres, $\alpha$ is the ratio of the post-collisional to the pre-collisional value of the normal component of their relative velocity (i.e.,  the component along the line joining the centers of the two spheres at contact). While the grains will be modeled as a gas of smooth inelastic hard spheres, the background gas will be effectively modeled as an interstitial fluid. Our goal will be the calculation of the mean square displacement (MSD)
of intruders or impurities (modeled as smooth inelastic hard spheres) immersed in this suspension. This endeavor is in line with what was done in
a previous work for a granular gas in the absence of any background fluid (dry granular gas)\cite{ABG22}.  In general, intruders and grains
will be assumed to be mechanically different, but we will also address the important special case of self-diffusion. 

An intruder can be viewed as part of a larger collection thereof, and in this sense one has a ternary mixture consisting of the intruders, the granular gas, and the molecular gas.
In this context, the present study is limited to a very low concentration of the intruders (tracer limit). In this extreme situation, one can assume that (i) the state of the granular gas (excess component) is not disturbed by the presence of the intruders and (ii) one can neglect the collisions between the intruders themselves in their kinetic equation. Nonetheless, the state of the intruders is determined by their interactions with the grains and the interstitial gas. 
Thus, while the velocity distribution function of the granular gas obeys the (nonlinear) Enskog kinetic equation, the velocity distribution function of the intruders satisfies the (linear) Enskog--Lorentz kinetic equation. Two different Fokker--Planck operators are incorporated into both kinetic equations to account for the influence of the background gas on grains and intruders, respectively.

In order to properly contextualize the present work, it is instructive to briefly revisit the previously addressed case of a \emph{dry} granular
gas. The MSD of intruders immersed in such gas has been recently determined in the HCS \cite{ABG22}. In this case, it is straightforward
to show that Haff's cooling law for the granular temperature in the HCS  \cite{H83} leads to a logarithmic time dependence of the intruder's
MSD (ultraslow diffusion). The drastic slowing down of diffusive transport comes as no surprise, since the continuous energy loss of the grains
is detrimental to the billiard-like motion of the intruder, which is deflected after each collision with a grain. The results derived in
Ref.\ \cite{ABG22} apply for arbitrary values of the masses of intruder and particles of the granular gas and hence, they extend the results derived in previous works devoted to the self-diffusion (intruders mechanically equivalent to granular gas particles)
\cite{BP00a,BChChM15} and the Brownian limit (intruder's mass much larger than the grain's mass) \cite{BRGD99}. An interesting
finding concerns the non-monotonic behavior of the MSD as a function of the coefficient of normal restitution $\alpha$ characterizing
the grain-grain collisions in the above system \cite{ABG22}. A random walk interpretation of the intruders' motion allows one to intuitively understand the observed $\alpha$ dependence of the MSD arising from the interplay between two competing effects  \cite{ABG22}; the increase of intruder-grain collision frequency with increasing  $\alpha$ on the one hand, and the concomitant decrease in the
persistence of the random walk on the other hand.

While the problem of tracer diffusion in the HCS is interesting from an academic point of view, it is not an easy task to
reproduce the conditions of the HCS in real situations. In addition, the theoretical and experimental study of the ultraslow diffusion
observed in the HCS
should be carried out with caution because of the memory effects and the (weak) ergodicity breaking implied by this type of stochastic
transport \cite{BChChM15}. In order to restore
\emph{normal} diffusion (linear time growth of the MSD), the energy loss of the intruder must be compensated for by an
external energy
supply, which can be modeled by a thermostat (in our case, the aforementioned interstitial fluid). The price to pay is that the
background fluid introduces a second source of dissipation in addition to the collisions between hard spheres, namely, viscous
drag forces respectively acting on the intruder and the grains. As already mentioned, the influence of the background molecular gas on
both intruders and grains will be modeled by a viscous drag force and by a stochastic Langevin-like force defined in terms of the
background (or bath) temperature $T_\text{b}$.

When the cooling effects arising from the viscous drag and from the dissipation due to inelastic collisions are exactly counterbalanced
by the energy gain of the grains provided by their interaction with the background gas, the system attains a steady state in which the
intruder exhibits normal diffusion. The corresponding diffusion coefficient $D$ displays a complex dependence on the
system parameters (masses, diameters, coefficients of normal restitution, density, and bath temperature). In particular, its dependence on
$\alpha$ leads to a non-monotonic behavior of the MSD, as it is the case in the HCS. We will show that a random walk picture
inspired by free path theory can again be used here to rationalize the observed behavior.

Like in our previous study for the HCS \cite{ABG22}, the determination of the tracer diffusion coefficient $D$ here is carried out by solving
the corresponding Enskog--Lorentz kinetic equation by means of the Chapman--Enskog method \cite{CC70} to first order in the
concentration gradient. A subtle point for granular suspensions is that the reference state of the perturbation scheme is a stationary distribution. Namely, the effect of the interstitial gas opens the possibility of a balanced energy transfer, and so the so-called homogeneous steady state is used to locally obtain the zeroth-order distribution function. As in the case of molecular gases \cite{CC70}, the coefficient $D$ is given by a linear integral equation that can be
solved by an expansion in Sonine polynomials. It turns out that  the diffusion coefficient satisfies a \emph{closed} equation, and it is thus not
coupled to the remaining transport coefficients. This important simplification with respect to the case of arbitrary concentration
\cite{GKG20} paves the way for obtaining the different Sonine approximations within a theoretical framework similar to that used for the HCS \cite{GM04,GV09,GV12}. Here, we go as far as the second Sonine approximation to determine $D$ in terms of the main mechanical
parameters of the suspension, i.e., masses and diameters of both species, the density, the (reduced) background temperature,
and the coefficients of restitution $\al_0$ and $\al$ for the intruder-grain collisions and the grain-grain collisions, respectively.
The present results for $D$ in the first-Sonine approximation are consistent with those recently obtained in Ref.\
\cite{GG22b}  in the low-density  regime. To test the accuracy of our theoretical results, we compare
them against numerical solutions of the Enskog--Lorentz equation performed by the direct simulation Monte Carlo (DSMC)
method \cite{B94}. As in previous works \cite{GM04,GV09,GV12}, the diffusion coefficient $D$ is readily extracted from the MSD of intruders provided by the simulations.

Apart from its academic interest, we think that our results can also be useful for understanding tracer diffusion in suspensions in some realistic situations. In particular, one of the main motivations of the present work has been to assess the relevance of grain-grain and grain-intruder collision effects when studying transport in granular suspensions. This aspect is relevant because collisions have not been considered in many of the previous works devoted to suspensions. Thus, in Sec.\ \ref{secAppli} we apply our theoretical results for describing a suspension of gold grains immersed in a hydrogen molecular gas. In view of the results given in Sec.\ \ref{secAppli}, it is quite apparent that the impact of collisions on the diffusion coefficient is not negligible for many of the situations in which the present suspension model applies.

The remainder of the paper is organized as follows. In Sec.\  \ref{sec2}, we study the steady state of the granular gas in thermal contact with a bath
(molecular gas) at temperature $T_\text{b}$. In particular, we determine the stationary granular temperature $T$ of the granular gas by
approximating its distribution function $f$ by a Maxwellian distribution. Even though this may seem a rough approximation, the obtained
dependence of $T$ on the coefficient of restitution $\al$ is in excellent agreement with computer simulations. The steady homogeneous state of the intruders plus granular gas in thermal contact with the molecular gas bath is studied in Sec.\ \ref{sec3}. As expected, the intruders' temperature $T_0$ is not the same as that of the granular gas $T$ ($T_0\neq T$); in other words, there is a breakdown of energy equipartition. In Sec.\ \ref{sec4}, the Chapman--Enskog method is applied
to solve the Enskog--Lorentz kinetic equation to first order in the concentration gradient. 
In Sec.\ \ref{sec5} we present Monte Carlo simulation results for the Enskog-Lorentz equation and compare the simulation data with the theoretical results for the temperatures $T$ and $T_0$ and intruder diffusion coefficient $D$ (obtained in the first and second
Sonine approximations).  Section \ref{sec6} is devoted to a comprehensive physical discussion of the diffusion properties of the intruder in terms of a random walk description. 
The applicability of the suspension model considered in this paper to real systems is discussed in Sec.\ \ref{secAppli}. Finally, a summary of the main results along with an outline of possible extensions of the present work is given  in Sec.\ \ref{sec7}. Some technical details
concerning the calculations of Sec.\ \ref{sec4} are provided in the Appendix \ref{appA}.

\section{Granular gas immersed in a molecular gas. Homogeneous state}
\label{sec2}

We consider a gas of solid particles modeled as smooth inelastic hard spheres of mass $m$ and diameter $\sigma$. The spheres (grains)
are immersed in a gas of viscosity $\eta_g$ and undergo instantaneous collisions between them.
As anticipated in Sec.\ \ref{sec1}, the inelasticity of collisions in the case of smooth spheres is fully characterized by the constant (positive)
coefficient of normal restitution $\al\leqslant 1$.


In the case of suspensions where the effect of the background gas on grain-grain collisions can be neglected  \cite{S20},   the influence of gas-phase effects on the dynamics of the solid particles is usually incorporated in an effective way via a fluid--solid interaction force in the starting kinetic equation \cite{K90,G94,J00}. Some models for granular suspensions \cite{LMJ91,TK95,SMTK96,WZLH09,PS12,H13,WGZS14,SA17,ASG19,SA20} only take into account
gas-solid interactions via Stokes'  linear drag force, which mimics the friction of grains with the interstitial gas. Here, we also
include an additional Langevin-like force  \cite{GTSH12} to account for the energy gained by the solid particles due to their interaction
with the background gas. Thus, for moderate densities (and assuming that the granular gas is in a steady homogeneous state), the one-particle
velocity distribution function $f(\mathbf{v},t)$ of the granular gas satisfies the nonlinear Enskog equation \cite{G19}
\beq
\label{2.1}
-\gamma\frac{\partial}{\partial\mathbf{v}}
\cdot\mathbf{v}f-\frac{\gamma T_{\text{b}}}{m}\frac{\partial^2 f}{\partial v^2}=J[\mathbf{v}|f,f],
\eeq
where the Enskog collision operator reads
\beqa
\label{2.2}
J\left[\mathbf{v}_1|f,f\right]&=&\chi \sigma^{d-1}\int d\mathbf{v}_2\int \dd\widehat{\boldsymbol{\sigma}}\Theta\left(\widehat{\boldsymbol{\sigma}}\cdot\mathbf{g}_{12}\right)
\left(\widehat{\boldsymbol{\sigma}}\cdot\mathbf{g}_{12}\right)\nonumber\\
& & \times
\Big[\alpha^{-2}f(\mathbf{v}_1'',t)f(\mathbf{v}_2'',t)-f(\mathbf{v}_1,t)f(,\mathbf{v}_2,t)\Big].
\nonumber\\
\eeqa
In the Enskog equation \eqref{2.1}, we have replaced $k_\text{B} T_\text{b}$ with $T_\text{b}$ for notational simplicity. This is equivalent to taking units of energy and/or temperature for which the Boltzmann constant $k_\text{B}$ is equal to 1. We will use this convention throughout this paper.  In addition, the symbol $\chi$ denotes the grain-grain pair correlation function at contact (i.e., when the distance between their centers is $\sigma$), $\widehat{\boldsymbol{\sigma}}$ is a unit vector directed along the line joining the centers of the colliding spheres, $\Theta$ stands for
the Heaviside step function [$\Theta(x)=1$ for $x> 0$, $\Theta(x)=0$ for $x\leq 0$], and $\mathbf{g}_{12}=\mathbf{v}_1-\mathbf{v}_2$ is the relative velocity of the two colliding spheres. The double primes on the velocities denote their precollisional values $(\mathbf{v}_1'',\mathbf{v}_2'')$ which yield the postcollisional values $(\mathbf{v}_1,\mathbf{v}_2)$:
\beq
\label{2.3}
\mathbf{v}_{1}''=\mathbf{v}_{1}- \frac{1+\alpha^{-1}}{2}\left(\boldsymbol{\widehat{\sigma}}
\cdot\mathbf{g}_{12}\right)\boldsymbol{\widehat{\sigma}},
\eeq
\beq
\label{2.3.1}
\mathbf{v}_2''=\mathbf{v}_2+\frac{1+\alpha^{-1}}{2}\left(\boldsymbol{\widehat{\sigma}}
\cdot\mathbf{g}_{12}\right)\boldsymbol{\widehat{\sigma}}.
\eeq
Equations \eqref{2.3} and \eqref{2.3.1} lead to the relation $\alpha \left(\boldsymbol{\widehat{\sigma}}\cdot\mathbf{g}_{12}''\right)=-\left(\boldsymbol{\widehat{\sigma}}
\cdot\mathbf{g}_{12}\right)$, where $\mathbf{g}_{12}''=\mathbf{v}_1''-\mathbf{v}_2''$.

In Eq.\ \eqref{2.1}, the amplitude of the stochastic force is selected to recover the fluctuation-dissipation theorem in the elastic limit. Here, $\gamma$ denotes the drift or friction coefficient
(characterizing the interaction between particles of the granular gas and the background gas), whereas $T_\text{b}$ stands for the bath
temperature. As in previous works \cite{GGG19a,GKG20}, we henceforth assume that $\gamma$ is a scalar quantity proportional to the
gas viscosity $\eta_g$ \cite{KH01}. In the \emph{dilute} limit, each particle is only subjected to its own Stokes drag. For hard
spheres ($d=3$), the corresponding drift coefficient is
\beq
\label{2.4.0}
\gamma\equiv \gamma_\text{St}=\frac{3 \pi \sigma \eta_g}{m}.
\eeq
For moderate densities and low Reynolds numbers, one has
\beq
\label{2.5}
\gamma=\gamma_\text{St}R(\phi),
\eeq
where $R(\phi)$ is a function of the solid volume fraction
\beq
\label{2.6}
\phi=\frac{\pi^{d/2}}{2^{d-1}d\Gamma \left(\frac{d}{2}\right)}n\sigma^d.
\eeq
The density dependence of the dimensionless function $R$ can be inferred from computer simulations. Specific forms of $R$ will be used
later to explore the dependence of dynamic system properties on various parameters. On the other hand, it is worth noting that the
main results  reported in this paper apply regardless of the specific choice for $R$.

It must be noted that the suspension model defined by Eq.\ \eqref{2.1} is a simplified version of the original Langevin-like model proposed in Ref.\ \cite{GTSH12}. In this latter model, the friction coefficient of the drag force ($\gamma$ in the notation of \cite{GTSH12}) and the strength of the correlation ($\xi$ in the notation of \cite{GTSH12}) are considered to be different in general. Instead, as done in several previous works on granular suspensions \cite{HTG17,GGG19a,THSG20}, we choose here to use the relation $\xi=2\gamma T_\text{b}/m$ for consistency with the fluctuation–dissipation theorem valid for elastic grain-grain collisions.

In the homogeneous state, the only nontrivial balance equation corresponds to the granular temperature $T$ defined as 
\beq
\label{2.5.0a}
  T=\frac{1}{n d} \int d\mathbf{v}\; m v^2 f(\mathbf{v}),
\eeq
where $n=\int d\mathbf{v}\; f(\mathbf{v})$ is the number density of solid particles.
The balance equation for $T$ can be easily derived by multiplying both sides of Eq.\ \eqref{2.1}
with $m v^2$ and subsequently integrating over velocity. One is then left with the result
\beq
\label{2.7}
2\gamma \left(T_\text{b}-T \right)=T \zeta,
\eeq
where
\beq
\label{2.8}
\zeta=-\frac{1}{n d T}\int d\mathbf{v}\; m v^2\; J[f,f]
\eeq
denotes the cooling rate, i.e., the energy loss per unit time due to collisions. Since $\zeta$ is a functional of the distribution $f(\mathbf{v})$, one needs to know $f$ to compute the cooling rate.


For inelastic collisions ($\al \neq 1$), dimensional analysis suggests the form $f(\mathbf{v})=n \pi^{-d/2} v_\text{th}^{-d} \varphi\left(v/v_\text{th}\right)$ where $v_\text{th}=\sqrt{2T/m}$ is the thermal velocity of the granular gas. So far, the exact form of the scaled distribution $\varphi$
is unknown, and one is therefore led to consider approximate forms for $f(\mathbf{v})$. In particular, previous results  \cite{GGG19a} show
that the stationary temperature $T$ is well approximated by a Maxwellian distribution:
\beq
\label{2.9.1}
f(\mathbf{v})\to f_\text{M}(\mathbf{v})=n\left(\frac{m}{2\pi T}\right)^{d/2}\exp\left(-\frac{m v^2}{2 T}\right).
\eeq
Making use of this approximation in the definition \eqref{2.8} of $\zeta$, one obtains the result 
\beq
\label{2.10}
\zeta=\frac{1-\al^2}{d}\nu,
\eeq
where the effective collision frequency $\nu$ is
\beq
\label{2.11}
\nu=\frac{\sqrt{2}\pi^{(d-1)/2}}{\Gamma\left(\frac{d}{2}\right)}n\sigma^{d-1}\chi v_\text{th}
=\frac{2^{d-1/2}d}{\sqrt{\pi}}\, \frac{\phi}{\sigma}\, \chi\,v_\text{th}.
\eeq
Even though one should bear in mind that the Maxwellian distribution \eqref{2.9.1} is not an exact solution of the Enskog equation \eqref{2.1}, it can be tentatively used to estimate the cooling rate $\zeta$ in the hope that the results will be sufficiently accurate for our purposes. As we will show in Sec.\ \ref{sec5}, the theoretical predictions for the granular temperature $T$ obtained with this Maxwellian approximation indeed yield an excellent agreement with computer simulations, which retrospectively justifies the use of this rather crude and yet successful approach.

Equation \eqref{2.10} allows one to rewrite Eq.\ \eqref{2.7} in dimensionless form as follows:
\beq
\label{2.12}
\lambda \left(T_\text{b}^*-T^*\right)=\zeta^* T^{*3/2},
\eeq
where $T_\text{b}^*=T_\text{b}/\mathcal{T}$, $T^*=T/\mathcal{T}$, $\zeta^*=\zeta/\nu=(1-\al^2)/d$, and 
\beq
\label{2.13}
\lambda=\frac{\sqrt{\pi}}{2^{d-1} d}\frac{R(\phi)}{\phi\chi}.
\eeq
Here, we have introduced the unit of energy
\beq
\label{2.13.1}
\mathcal{T}=m\sigma^2 \gamma_\text{St}^2.
\eeq
Equation \eqref{2.12} is a cubic equation for the (reduced) temperature $T^*$. The physical solution provides an expression for $T^*$ in
terms of  $\alpha$, $\phi$, and $T_\text{b}^*$ which must of course satisfy the requirement $\al=1$, $T^*=1$ for any value of $\phi$, and
$T_\text{b}^*$; in contrast, $T^*<T_\text{b}^*$ for any $\alpha<1$. More explicitly, Eq.\ \eqref{2.12} can be cast in the simplified form $\xi x^3+x^2-1=0$, 
with $x=\sqrt{T^*/T_\text{b}^*}$ and 
$\xi=\zeta^*\sqrt{T_\text{b}^*}/\lambda$.
The physical solution $x_\text{phys}$ must  correctly reproduce the elastic limit, i.e., $x_\text{phys}\to 1$ as $\alpha \to 1$. This yields 
\beq
\label{2.15}
x_\text{phys}=
\frac{\mathcal{X}^{1/3}+
\mathcal{X}^{-1/3}-1}{3\xi}
\eeq
with 
\beq
\mathcal{X}=
\frac{3\sqrt{3}\sqrt{27\xi^4-4\xi^2}+27\xi^2-2}{2}.
\eeq

Equation \eqref{2.15} can now be used as a starting point to devise a number of approximations for transport quantities of interest. In particular, one may consider the quasielastic regime $\alpha\lesssim 1$. We will return to this point when addressing the behavior of the intruder's diffusion coefficient in the self-diffusion case (cf.\ Secs. IV.B and V.B).

\section{Intruders in granular suspensions}
\label{sec3}


Let us now assume that some impurities, i.e., intruders of mass $m_0$ and diameter $\sigma_0$ are added to the granular gas (the zero subscripts will
hereafter denote quantities referred to such intruders). As already mentioned, the concentration of the intruders will be assumed to be negligibly small, implying that the number density of intruders $n_0$ [defined below in Eq.\ \eqref{3.6.2}] is much smaller than its counterpart $n$ for the grains (particles of the granular gas).  
Formally, the resulting system can be regarded as a binary granular mixture in which
one of the components is present in tracer concentration. For conciseness, in the remainder we will speak of intruders immersed in a granular gas instead of a binary granular mixture with one tracer component.

Further, we assume that both intruders and grains are
surrounded by an interstitial fluid. As in the case of the granular gas, we also assume that the surrounding fluid has no explicit influence on the intruder-grains collision rules. The inelasticity of these binary collisions is characterized by the coefficient of normal restitution $\al_0$, with $\al_0\neq \al$ in general. We recall our assumption of a very low intruder
concentration with negligible impact on the state of the granular gas. The intruder's interaction with the interstitial fluid is characterized by
the friction coefficient $\gamma_0$, which is in general different from $\gamma$. 

In the tracer limit, the intruder's velocity distribution function $f_0(\mathbf{r}, \mathbf{v}; t)$ obeys the Enskog equation
\beq
\label{3.1}
\frac{\partial f_0}{\partial t}+\mathbf{v}\cdot \nabla f_0-\gamma_0\frac{\partial}{\partial\mathbf{v}}\cdot\mathbf{v}f_0-\frac{\gamma_0 T_{\text{b}}}{m_0}\frac{\partial^2 f_0}{\partial v^2}=J_0[\mathbf{r}, \mathbf{v}|f_0,f],
\eeq
where the Enskog--Lorentz collision operator $J_0[f_0,f]$ reads \cite{G19}
\beqa
\label{3.2}
J_{0}\left[\mathbf{r}_1, \mathbf{v}_1|f_0,f\right]&=&\overline{\sigma}^{d-1}\int \dd\mathbf{v}_2\int \dd\widehat{\boldsymbol{\sigma}}\Theta\left(\widehat{\boldsymbol{\sigma}}\cdot\mathbf{g}_{12}\right)
\left(\widehat{\boldsymbol{\sigma}}\cdot\mathbf{g}_{12}\right)\nonumber\\
& & \times \Big[\alpha_{0}^{-2}\chi_{0}
(\mathbf{r}_1,\mathbf{r}_1-\boldsymbol{\overline{\sigma}})f_0(\mathbf{r}_1,\mathbf{v}_1'',t)
\nonumber\\
& &
\times f(\mathbf{v}_2'',t)
-\chi_{0}(\mathbf{r}_1,\mathbf{r}_1+\boldsymbol{\overline{\sigma}})f_0(\mathbf{r}_1,\mathbf{v}_1,t)
\nonumber\\
& & \times
f(\mathbf{v}_2,t)\Big].
\eeqa
Here, $\chi_0$ stands for the intruder-grain pair correlation function at contact , i.e., when the distance between their centers is $\overline{\sigma}=(\sigma+\sigma_0)/2$). Besides, 
 $\boldsymbol{\overline{\sigma}}=\overline{\sigma} \widehat{\boldsymbol{\sigma}}$  and $\widehat{\boldsymbol{\sigma}}$ is the unit vector
along the line joining the centers of the spheres that represent the intruder and the grain at contact. The precollisional velocities $(\mathbf{v}_1'',\mathbf{v}_2'')$ and their postcollisional counterparts $(\mathbf{v}_1,\mathbf{v}_2)$ are related to one another as
follows:
\beq
\label{3.2.1}
\mathbf{v}_1''=\mathbf{v}_1-\mu \left(1+\alpha_0^{-1}\right)\left(\boldsymbol{\widehat{\sigma}}
\cdot\mathbf{g}_{12}\right)\boldsymbol{\widehat{\sigma}},
\eeq
\beq
\label{3.2.2}
\mathbf{v}_2''=\mathbf{v}_2+\mu_0 \left(1+\alpha_0^{-1}\right)\left(\boldsymbol{\widehat{\sigma}}
\cdot\mathbf{g}_{12}\right)\boldsymbol{\widehat{\sigma}},
\eeq
where 
\beq
\label{3.2.2.0}
\mu=\frac{m}{m+m_0}, \quad  \mu_0=\frac{m_0}{m+m_0}. 
\eeq
From Eqs.\ \eqref{3.2.1} and \eqref{3.2.2}, one achieves the relation $\alpha_0 \left(\boldsymbol{\widehat{\sigma}}\cdot\mathbf{g}_{12}''\right)=-\left(\boldsymbol{\widehat{\sigma}}
\cdot\mathbf{g}_{12}\right)$.

Equations \eqref{3.2.1} and \eqref{3.2.2} correspond to the so-called inverse or \emph{restituting} collisions. Inversion of these collision rules yield the so-called \emph{direct} collisions, where
the precollisional velocities $(\mathbf{v}_1,\mathbf{v}_2)$ yield $(\mathbf{v}_1',\mathbf{v}_2')$ as post-collisional velocities:
\beq
\label{3.2.3}
\mathbf{v}_1'=\mathbf{v}_1-\mu \left(1+\alpha_0\right)\left(\boldsymbol{\widehat{\sigma}}
\cdot\mathbf{g}_{12}\right)\boldsymbol{\widehat{\sigma}},
\eeq
\beq
\label{3.2.4}
\mathbf{v}_2'=\mathbf{v}_2+\mu_0 \left(1+\alpha_0\right)\left(\boldsymbol{\widehat{\sigma}}
\cdot\mathbf{g}_{12}\right)\boldsymbol{\widehat{\sigma}}.
\eeq
Equations \eqref{3.2.3} and \eqref{3.2.4} yield the relation $\left(\boldsymbol{\widehat{\sigma}}\cdot\mathbf{g}_{12}'\right)=-\alpha_0\left(\boldsymbol{\widehat{\sigma}}
\cdot\mathbf{g}_{12}\right)$, where $\mathbf{g}_{12}'=\mathbf{v}_1'-\mathbf{v}_2'$.

Note also that Eq.\ \eqref{3.2} takes into account that the granular gas is in a homogeneous state, as there is no
dependence of $f$ on position. Moreover, if the distribution function of the granular gas $f$ were known, Eq.\ \eqref{3.1} would immediately become a linear equation for the distribution function  $f_0$ of the intruder velocities. However, as mentioned in Sec.\ \ref{sec2}, the distribution $f$ is yet to be found to date.

In accordance with Eq.\ \eqref{2.5}, the friction coefficient $\gamma_0$ for the intruder
in the regime of low Reynolds number takes the form
\beq
\label{3.3}
\gamma_0=\gamma_{0,\text{St}}R_0,
\eeq
where, for $d=3$,
\beq
\label{3.4}
\gamma_{0,\text{St}}=\frac{3 \pi \sigma_0 \eta_g}{m_0}=\frac{\sigma_0 m}{\sigma m_0}\gamma_{\text{St}}.
\eeq
As in the case of $R$, the dependence of the function $R_0$ on the density $\phi$ and on the remaining system
parameters will be borrowed from computer simulations (see Sec.\ \ref{sec5}).

\subsection{Homogeneous steady state}

In the absence of spatial gradients, the kinetic equation \ \eqref{3.1} becomes stationary and homogeneous in the long-time limit, i.e.,
\beq
\label{3.5}
-\gamma_0\frac{\partial}{\partial\mathbf{v}}
\cdot\mathbf{v}f_0-\frac{\gamma_0 T_{\text{b}}}{m_0}\frac{\partial^2 f_0}{\partial v^2}=J_0[f_0,f].
\eeq
From this equation, one easily finds that the stationary granular temperature of the intruder, defined as
\beq
\label{2.5.0}
T_0=\frac{1}{n_0 d} \int d\mathbf{v}\; m_0 v^2 f_0(\mathbf{v}),
\eeq
satisfies the relation
\beq
\label{3.6}
2\gamma_0 \left(T_\text{b}-T_0\right)=T_0 \zeta_0.
\eeq
Here,
\beq
\label{3.5.1}
\zeta_0=-\frac{1}{d n_0 T_0}\int d\mathbf{v}\; m_0 v^2\; J_0[f_0,f]
\eeq
is the partial cooling rate characterizing the rate of dissipation due to intruder-grain collisions. As in the case of the granular gas,
for elastic collisions ($\al_0=\al=1$), $\zeta_0=0$, $T_\text{b}=T_0$, and so Eq.\ \eqref{3.5} has the exact solution
\beq
\label{3.6.1}
f_0(\mathbf{v})=n_0 \left(\frac{m_0}{2\pi T_\text{b}}\right)^{d/2} \exp \left(-\frac{m_0 v^2}{2 T_\text{b}}\right),
\eeq
where
\beq
\label{3.6.2}
n_0(\mathbf{r};t)=\int d\mathbf{v}\; f_0(\mathbf{r}, \mathbf{v}; t)
\eeq
is the number density of intruders. 

For inelastic collisions ($\al \neq 1$), we already mentioned that $f$ is
not known, implying that the solution of Eq.\ \eqref{3.5} and the exact form of the cooling rate \eqref{3.5.1} are also unknown.
However, as in the case of the cooling rate $\zeta$, a good estimate for $\zeta_0$ can be obtained by respectively replacing $f$ and $f_0$ with Maxwellian distributions defined at the
temperatures $T$ and $T_0$, respectively \cite{GGG21}. The Maxwellian approximation for $f$ is given by \eqref{2.9.1}; similarly, for $f_0$
one performs the replacement
\beq
\label{3.7}
f_0(\mathbf{v})\to f_{0,\text{M}}(\mathbf{v})=n_0\left(\frac{m_0}{2\pi T_0}\right)^{d/2}\exp\left(-\frac{m_0 v^2}{2 T_0}\right).
\eeq
With this approach, the (reduced) partial cooling rate $\zeta_0^*=\zeta_0/\nu$ takes the form \cite{G19}
\beqa
\label{3.8}
\zeta_0^*&=&\frac{2\sqrt{2}}{d} \mu \frac{\chi_0}{\chi} \left(\frac{\overline{\sigma}}{\sigma}\right)^{d-1} \left(\frac{1+\beta}{\beta}\right)^{1/2}(1+\al_{0})\nonumber\\
& & \times\left[1-\frac{1}{2}\mu (1+\beta)(1+\al_{0})\right],
\eeqa
where
\beq
\label{3.8.1}
\beta=\frac{m_0 T}{m T_0}
\eeq
denotes the ratio between the mean square velocities of intruders and grains.

In dimensionless form, Eq.\ \eqref{3.6} can be rewritten as
\beq
\label{3.9}
\lambda_0 \left(T_\text{b}^*-T_0^*\right)=\zeta_0^* \sqrt{T^{*}}T_0^*,
\eeq
where $T_0^*=T_0/\mathcal{T}$ and
\beq
\label{3.10}
\lambda_0=\frac{\sqrt{\pi}}{2^{d-1} d}\frac{\gamma_{0,\text{St}}}{\gamma_{\text{St}}}\frac{R_0}{\phi\chi}.
\eeq
When intruder and granular gas particles are mechanically equivalent ($m=m_0$, $\sigma=\sigma_0$, and $\al=\al_0$),
then $\lambda=\lambda_0$, $\zeta^*=\zeta_0^*$, $T^*=T_0^*$, and hence energy equipartition applies. In the
general case (namely, when collisions are inelastic, and intruder and grains are mechanically different), the solution to the
cubic equation \eqref{3.9} provides $T_0^*$ in terms of the system parameters. As in the freely cooling case \cite{G19}, one finds that there is a breakdown of energy equipartition ($T_0^*\neq T^*$) as expected.

\section{Tracer diffusion coefficient}
\label{sec4}

We now turn to our main goal, namely, the computation of the diffusion coefficient of intruders immersed in a granular suspension. We
make the usual assumption that the diffusion process is triggered by the presence of a weak concentration gradient $\nabla n_0$,
which for simplicity is taken to be the only gradient in the system. In this setting, the kinetic equation for $f_0$ is given by Eq.\ \eqref{3.1},
where one takes $\chi_0\equiv \text{const.}$ in the Enskog-Lorentz collision operator $J_0[f_0,f]$. Since intruders may freely
exchange momentum and energy with the grains, only their number density $n_0$ is conserved:
\beq
\label{4.1}
\frac{\partial n_0}{\partial t}=-\nabla \cdot \mathbf{j}_0,
\eeq
where
\beq
\label{4.2}
\mathbf{j}_0(\mathbf{r}; t)= \int d\mathbf{v}\; \mathbf{v}\; f_0(\mathbf{r}, \mathbf{v};t)
\eeq
is the intruder particle flux.

Equation \eqref{4.1} becomes a closed hydrodynamic equation for the hydrodynamic field $n_0$ once the flux $\mathbf{j}_0$ is expressed as a functional of $n_0$ and $T$. Our aim here is to obtain the intruder particle flux to first order in $\nabla n_0$ by applying the Chapman--Enskog method \cite{CC70} adapted to dissipative dynamics \cite{G19}. Thus, at times much longer than the mean free time, we expect the system
to reach a hydrodynamic regime in which the Enskog equation \eqref{3.1} admits a \emph{normal} solution. This means that all the space and
time dependence only enters $f_0$ via its functional dependence on the hydrodynamic fields $n_0$ and $T$, i.e.,
\beq
\label{4.3}
f_0(\mathbf{r}, \mathbf{v};t)=f_0\left[n_0(\mathbf{r};t), T(t)\right].
\eeq
In writing \eqref{4.3}, we have assumed that the distribution $f$ for the granular gas also adopts the normal form.
Assuming a small strength of the density gradient, one can express $f_0$ in terms of an expansion of powers of $\nabla n_0$:
\beq
\label{4.4}
f_0=f_0^{(0)}+\epsilon f_0^{(1)}+\epsilon^2 f_0^{(2)}+\cdots,
\eeq
where $\epsilon$ is a formal parameter measuring the non-uniformity of the system; in fact, here each factor $\epsilon$ has the implicit
meaning of a  $\nabla n_0$ factor. In this paper, only terms to first-order in $\epsilon$ will be considered.

The time derivative $\partial_t$ is also expanded as $\partial_t=\partial_t^{(0)}+\epsilon \partial_t^{(1)}+\cdots$, where
\beq
\label{4.5}
\partial_t^{(0)}n_0=0, \quad \partial_t^{(0)}T=2\gamma\left(T_\text{b}-T\right)-\zeta T,
\eeq
\beq
\label{4.6}
\partial_t^{(1)} n_0=-\nabla \cdot \mathbf{j}_0^{(0)}, \quad \partial_t^{(1)}T=0,
\eeq
with
\beq
\label{4.6.1}
\mathbf{j}_0^{(0)}= \int d\mathbf{v}\; \mathbf{v}\; f_0^{(0)}(\mathbf{v};t).
\eeq

As noted in previous works \cite{GChV13a,GKG20}, although we are interested in computing the diffusion
coefficient in the steady-state, the presence of the interstitial fluid
introduces the possibility of a local energy imbalance, and,
hence, the zeroth-order distribution $f_0^{(0)}$  is in general a time-dependent one.
This is because, for arbitrarily small deviations
from the homogeneous steady state, the energy gained by grains through
collisions with the background fluid cannot be
locally compensated for by the cooling terms arising from
viscous friction and collisional dissipation. Thus, in
order to obtain the tracer diffusion coefficient $D$ in the steady state, one first has
to determine the \emph{time-dependent} integral equation satisfied by this quantity,
and then solve this equation under the steady-state condition \eqref{2.7}.

In the hydrodynamic regime \cite{CC70}, the zeroth-order approximation $f_0^{(0)}$ only depends on time via the granular temperature $T$. In this case,  $\partial_t^{(0)}f_0^{(0)}=(\partial_T f_0^{(0)})(\partial_t^{(0)}T)$ and hence, 
the distribution $f_0^{(0)}$ satisfies the kinetic equation
\beq
\label{4.7}
T \Delta \frac{\partial f_0^{(0)}}{\partial T}-\gamma_0\frac{\partial}{\partial\mathbf{v}}\cdot\mathbf{v}f_0^{(0)}-\frac{\gamma_0 T_{\text{b}}}{m_0}\frac{\partial^2 f_0^{(0)}}{\partial v^2}=J_0[f_0^{(0)},f],
\eeq
where
\beq
\label{4.7.1}
\Delta\equiv 2\gamma\left(\frac{T_\text{b}}{T}-1\right)-\zeta.
\eeq
In the steady state ($\Delta=0$), Eq.\ \eqref{4.7} takes the same form as Eq.\ \eqref{3.5}, except that the zeroth-order solution
$f_0^{(0)}(\mathbf{r}, \mathbf{v};t)$ is now a \emph{local} distribution function. The stationary solution of Eq.\ \eqref{4.7} has already
been discussed in Sec.\ \ref{sec3}. Since $f_0^{(0)}(\mathbf{v})$ is isotropic in $\mathbf{v}$, then $\mathbf{j}_0^{(0)}=\mathbf{0}$. Thus, according to Eq.\ \eqref{4.6}, $\partial_t^{(1)} n_0=0$.

To first-order in $\nabla n_0$, one obtains the kinetic equation
\beq
\label{4.8}
-\gamma_0\frac{\partial}{\partial\mathbf{v}}\cdot\mathbf{v}f_0^{(1)}-\frac{\gamma_0 T_{\text{b}}}{m_0}\frac{\partial^2 f_0^{(1)}}{\partial v^2}-J_0[f_0^{(1)},f]=-\frac{f_0^{(0)}}{n_0} \mathbf{v}\cdot \nabla n_0.
\eeq
To derive Eq.\ \eqref{4.8}, we have assumed stationarity ($\Delta=0$)
and have taken into account that $\nabla f_0^{(0)}=f_0^{(0)}\nabla \ln n_0$. The solution to Eq.\ \eqref{4.8} is proportional to $\nabla n_0$:
\beq
\label{4.9}
f_0^{(1)}(\mathbf{v})=\boldsymbol{\mathcal A}(\mathbf{v}) \cdot \nabla n_0,
\eeq
where the coefficient $\boldsymbol{\mathcal A}$ is a function of the velocity and the hydrodynamic fields. To first order of $\nabla n_0$,
the intruder flux reads
\beq
\label{4.10}
\mathbf{j}_0^{(1)}=\int d\mathbf{v}\; \mathbf{v}\; f_0^{(1)}(\mathbf{v})= -D \nabla n_0,
\eeq
where $D$ stands for the diffusion coefficient. Use of Eq.\ \eqref{4.9} in Eq.\ \eqref{4.10} allows one to define the coefficient $D$ as
\beq
\label{4.11}
D=-\frac{1}{d}\int d\mathbf{v}\; \mathbf{v}\cdot \boldsymbol{\mathcal A}(\mathbf{v}).
\eeq
Substitution of Eq.\ \eqref{4.9} into Eq.\ \eqref{4.8} yields the following linear integral equation for the unknown $\boldsymbol{\mathcal A}$:
\beq
\label{4.11.1}
-\gamma_0\frac{\partial}{\partial\mathbf{v}}\cdot\mathbf{v}\boldsymbol{\mathcal A}-\frac{\gamma_0 T_{\text{b}}}{m_0}\frac{\partial^2 \boldsymbol{\mathcal A}}{\partial v^2}-J_0[\boldsymbol{\mathcal A},f]=-\frac{f_0^{(0)}}{n_0} \mathbf{v}.
\eeq

Let us now write the diffusion equation. Substitution of Eq.\ \eqref{4.10} into Eq.\ \eqref{4.1} leads to
\beq
\label{4.28}
\frac{\partial n_0}{\partial t}=D \nabla^2 n_0.
\eeq
As for elastic collisions, Eq.\ \eqref{4.28} is a diffusion equation with a diffusion coefficient $D$ that is constant in time.
Thus, we can immediately write the intruder's MSD at time $t$ as
\beq
\label{4.29}
\langle |\Delta \mathbf{r}|^2(t)\rangle=2 d D t.
\eeq
Equation \eqref{4.29} is the Einstein form, relating the diffusion coefficient to the MSD. The relation \eqref{4.29} will be used in Monte Carlo simulations of granular gases to measure the diffusion coefficient.

\subsection{First and second Sonine approximations to $D$}

Equation \eqref{4.11} describes the dependence of the diffusion coefficient on $\boldsymbol{\mathcal A}(\mathbf{v})$, which is in turn given
by the solution of the integral equation \eqref{4.11.1}. This equation can be approximately solved by using a Sonine polynomial expansion.
This expansion can be truncated to different orders, resulting in increasingly accurate approximations. As mentioned in Sec.\ \ref{sec1},  we will restrict ourselves to the first and the second-order to compute $D$, i.e., to the so-called first and second Sonine approximations. Up to the second
Sonine approximation, $\boldsymbol{\mathcal A}(\mathbf{v})$ is estimated by the expression
\beq
\label{4.12}
\boldsymbol{\mathcal A}(\mathbf{v})\to -f_{0,\text{M}}(\mathbf{v})\Big[a_1\mathbf{v}+a_2 \mathbf{S}_0(\mathbf{v})\Big],
\eeq
where $f_{0,\text{M}}(\mathbf{v})$ is defined by Eq.\ \eqref{3.7}, and $\mathbf{S}_0(\mathbf{v})$ is the polynomial
\beq
\label{4.13}
\mathbf{S}_0(\mathbf{v})=\Big(\frac{1}{2}m_0 v^2-\frac{d+2}{2}T_0\Big)\mathbf{v}.
\eeq
The Sonine coefficients $a_1$ and $a_2$ are defined as
\beq
\label{4.14}
a_1=-\frac{m_0}{d n_0 T_0}\int d\mathbf{v}\; \mathbf{v}\cdot \boldsymbol{\mathcal A}(\mathbf{v})=\frac{m_0 D}{n_0 T_0},
\eeq
\beq
\label{4.15}
a_2=-\frac{2}{d(d+2)}\frac{m_0}{n_0 T_0^3}\int d\mathbf{v}\; \mathbf{S}_0(\mathbf{v})\cdot \boldsymbol{\mathcal A}(\mathbf{v}).
\eeq
The evaluation of the coefficients $a_1$ and $a_2$ is carried out in the Appendix \ref{appA}.

\begin{figure}
\begin{center}
\includegraphics[width=.7\columnwidth]{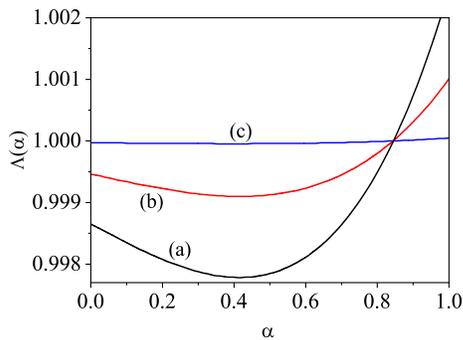}
\end{center}
\caption{Plot of the function $\Lambda=D^*[2]/D^*[1]$ vs the (common) coefficient of restitution $\al=\al_0$ for three different systems: (a) $m_0/m=1/8$ and $\sigma_0/\sigma=1/2$; (b) $m_0/m=(0.2)^3$ and $\sigma_0/\sigma=0.2$; and (c) $m_0/m=8$ and $\sigma_0/\sigma=2$.
\label{fig1}}
\end{figure}


The tracer diffusion coefficient $D$ can be written in terms of a \emph{reduced} diffusion coefficient $D^*$ as
\beq
\label{4.16}
D=\gamma_{\text{St}}\sigma^2 D^*,
\eeq
which, in fact, implies the use of $\gamma_{\text{St}}^{-1}$ and $\sigma$ as time and length units, respectively. The advantage of choosing $\gamma_{\text{St}}$ as time unit instead of the effective collision frequency $\nu(T)\propto \sqrt{T(\al)}$ is that the former does not
depend on the coefficient of restitution $\al$. The expression for $D^*$ depends on the Sonine
approximation considered. In particular, the second Sonine approximation $D^*[2]$ to $D^*$ gives
\beq
\label{4.17}
D^*[2]=\Lambda D^*[1],
\eeq
where
\beq
\label{4.18}
\Lambda=\frac{(\nu_a^*+\gamma_0^*)(\nu_d^*+3\gamma_0^*)}{(\nu_a^*+\gamma_0^*)(\nu_d^*
+3\gamma_0^*)-\nu_b^*\Big[\nu_c^*+2\gamma_0^*
\Big(1-\frac{T_\text{b}^*}{T_0^*}\Big)\Big]}.
\eeq
The (reduced) collision frequencies $\nu_a^*$, $\nu_b^*$, $\nu_c^*$, and $\nu_d^*$ are given in the Appendix \ref{appA}. In Eq.\ \eqref{4.17}, $D^*[1]$ denotes the first Sonine approximation for $D^*$, which reads
\beq
\label{4.19}
D^*[1]=\frac{m}{m_0}\frac{T_0^*}{R}\frac{\gamma^*}{\gamma_0^*+\nu_a^*},
\eeq
where
\beq
\label{4.20}
\gamma^*=\frac{\gamma}{\nu}=\frac{\sqrt{\pi}}{2^{d} d}\frac{R}{\phi\chi\sqrt{T^*}}, \quad \gamma_0^*=\frac{\gamma_0}{\nu}=\frac{\gamma_{0,\text{St}}}{\gamma_{\text{St}}} \frac{R_0}{R}\gamma^*.
\eeq
The expression \eqref{4.19} of $D^*[1]$ is consistent with the one derived in Ref.\ \cite{GKG20} for arbitrary concentration and, more recently, with that derived in Ref.\ \cite{GG22b} in the low-density regime. 

To illustrate the discrepancy between the first and the second Sonine approximation for different parameter values,  we consider a binary
mixture with identical mass densities of intruders and grains, $m_0/m=(\sigma_0/\sigma)^3$ for $d=3$. Figure \ref{fig1} shows
the dependence of $\Lambda=D^*[2]/D^*[1]$ on the (common) coefficient of restitution $\al=\al_0$ for $d=3$, $T_\text{b}^*=1$, $\phi=0.1$, and three different values of the mass and diameter ratios. As it is the case for \emph{dry} granular gases (absence of interstitial gas) \cite{GM04}, the
departure of $\Lambda$ from unity is more significant when the intruders become much lighter than the grains. On the other
hand, one also sees that the correction of the second Sonine approximation to the first one is much weaker here than in the absence
of the interstitial gas. Thus, one expect that the first Sonine approximation will already come quite close to the exact value of the diffusion
coefficient, even for mass and/or diameter ratios that are far from unity. This point will be confirmed later when we compare
the theoretical predictions of both $D^*[2]$ and $D^*[1]$ with computer simulations.

As it turns out,  in general the coefficients $D^*[1]$ and $D^*[2]$ exhibit a complicated dependence on the system parameters,
i.e., the coefficients of restitution $\al$ and $\al_0$, the mass ratio $m_0/m$, the diameter ratio $\sigma_0/\sigma$, the density $\phi$,
and the (reduced) background temperature $T_\text{b}^*$. Moreover, our analytical results are valid for arbitrary Euclidean dimension $d$.
Before studying this dependence in detail, it is instructive to consider some special limiting cases. For simplicity, we will restrict the analysis
of these limiting cases to the first Sonine approximation $D^*[1]$ to $D^*$.

\subsection{Self-diffusion case}

By definition, in the self-diffusion limit case the intruder becomes indistinguishable from the grains as far as its mechanical properties are
concerned, i.e., $m=m_0$, $\sigma=\sigma_0$, $\al=\al_0$. In this limiting case, $T^*=T_0^*$, $\nu_a^*=(1+\al)/d$, and
\beq
\label{4.21}
\gamma^*=\gamma_0^*=\frac{\sqrt{\pi}}{2^{d} d}\frac{R}{\phi\chi\sqrt{T^*}}.
\eeq
Hence, the self-diffusion coefficient $D^*[1]$ can be written as
\beq
\label{4.22}
D^*[1]=\frac{T^*}{R+\frac{2^d}{\sqrt{\pi}} \phi \chi \sqrt{T^*}(1+\al)}.
\eeq
Let us now suppose that the force exerted by the interstitial gas is much greater than the effect of the collisions. This situation is formally equivalent to considering a very large (reduced) friction parameter, i.e. 
\beq
\label{lowstokes1}
\gamma^*=\gamma_0^*\gg 1\implies \gamma=\gamma_0\gg \nu.
\eeq
In the context of granular suspensions, Eq.\ \eqref{lowstokes1} implies low-Stokes numbers ($\text{St}\propto 1/ {\gamma^*}\to 0$). In this limit, $\gamma^*\gg \nu_a^*$ and Eq.\ \eqref{4.19} yields $D^*[1]=T^*$ in the self-diffusion case for a very dilute suspension ($R=1$). According to Eq.\ \eqref{2.7}, $T=T_\text{b}$ and so one recovers the standard Stokes--Einstein equation:
\begin{equation}
\label{DStoEin}
D_\text{SE}=\frac{T_\text{b}}{3\pi\sigma \eta_g}.
\end{equation}
We have used  Eqs.~\eqref{2.4.0}  and \eqref{4.16} in the derivation of Eq.\ \eqref{DStoEin}. An alternative way to achieve a low-Stokes number is to consider that $\phi\to 0$, but keeping $\sigma\equiv \text{finite}$. Notice that this is not the so-called Boltzmann--Grad limit ($\phi/\sigma\equiv \text{finite}$) on which the Boltzmann kinetic equation is based \cite{C90a, C94}. When the volume fraction tends to zero ($\phi\to 0$), $R(0)=1$ [see Eq.\ \eqref{5.3}] and according to Eq.\ \eqref{4.22} $D^*[1]=T^*$. The Stokes--Einstein equation is then recovered by making use of Eq.~\eqref{4.16} and the fact that $T\to T_\text{b}$ as $\phi\to 0$
[see Eqs.~\eqref{2.12} and \eqref{2.13}]. The lack of $\al$-dependence of the self-diffusion coefficient $D$ in the $\phi \to 0$ limit with $\sigma$ finite is indeed an expected result, since the mean-free time between collisions is much greater than the time taken by the gas-solid forces to significantly affect the motion of grains. Consequently, the inelasticity of such collisions becomes
increasingly irrelevant, and the diffusion of grains is only influenced by their interaction with the interstitial gas.

\subsection{Brownian diffusion case}

The Brownian diffusion regime is characterized by the conditions $m_0/m \to \infty$. In this limit case, Eqs. \eqref{3.8} and \eqref{3.10} respectively yield $\zeta_0^*\to (m/m_0) \widetilde{\zeta}_0$ and $\lambda_0\to (m/m_0) \widetilde{\lambda}_0$, where
\beq
\label{4.23}
\widetilde{\zeta}_0=\frac{2\sqrt{2}}{d}\frac{\chi_0}{\chi} \left(\frac{\overline{\sigma}}{\sigma}\right)^{d-1}(1+\al_0)\left[1-\frac{T^*}{2T_0^*}(1+\al_0)\right],
\eeq
and
\beq
\label{4.24}
\widetilde{\lambda}_0=\frac{\sqrt{\pi}}{12}\frac{\sigma_0}{\sigma}\frac{R_0}{\phi\chi}.
\eeq
To write Eq.\ \eqref{4.24}, use has been made of the relation \eqref{3.4} for $d=3$. Hence, according to Eq.\ \eqref{3.9}, $T_0^*$ can be written as
\beq
\label{4.25}
T_0^*=\frac{T_\text{b}^*+\left(\frac{1+\al_0}{2}\right)^2\kappa_g T^{*}\sqrt{T^*}}{1+\frac{1+\al_0}{2}\kappa_g \sqrt{T^*}},
\eeq
where
\beq
\label{4.26}
\kappa_g=\frac{2^{\frac{9}{2}}}{\sqrt{\pi}}\left(\frac{\overline{\sigma}}{\sigma}\right)^{2}\frac{\sigma}{\sigma_0}
\frac{\phi\chi_0}{R_0}.
\eeq
Moreover, it is straightforward to prove that, in the Brownian limit, the scaled diffusion coefficient $D^*[1]$ is
\beq
\label{4.27}
D^*[1]=\frac{\sigma}{\sigma_0 R_0}\frac{T_0^*}{1+\frac{1+\al_0}{2}\kappa_g\sqrt{T^*}}.
\eeq
The expression \eqref{4.27} agrees with the one obtained in a previous study of granular Brownian motion
by Sarracino \emph{et al.} \cite{SVCP10} based on a  model with $\gamma=\gamma_0\equiv\text{const}$.

In the same way as before, in the limit  $\phi\to 0$ but  $\sigma\equiv \text{finite}$, $R_0=1$ [see Eq.\ \eqref{5.5}], $\kappa_g\to 0$, and Eq.\ \eqref{4.27} leads to $D[1]=T_0/(3\pi \sigma_0 \eta_g)$. As in the self-diffusion case, the diffusion coefficient is given by the standard Stokes--Einstein equation \eqref{DStoEin}, as can be shown from the fact that $T_0\to T \to T_\text{b}$ as $\phi\to 0$. The physical justification of this result follows the same lines as in the self-diffusion case: intruder-grain collisions become increasingly rare,
and the dynamics of intruders are essentially driven by their interaction with the bath.

\section{Comparison between theory and Monte Carlo simulations}
\label{sec5}

\subsection{Simulation method}

In this section we compare the theoretical predictions of Sec.\ \ref{sec4} with the results obtained by numerically solving the Enskog equation
by means of the DSMC method. The adaptation of DSMC method to study binary granular suspensions has
been described in some detail in the literature (see, e.g., Refs.\ \cite{GKG21,GGG21}). Here, we only mention some specifities of the
tracer limit ($n_0/n\to 0$), which justifies the use of the term ``intruder''. Due to the very low intruder concentration, intruder-intruder
collisions are rare, and so their effect will be neglected.  Besides, when an intruder collides with a grain, the post-collisional
velocity obtained from the scattering rules \eqref{3.2.3} and \eqref{3.2.4} is only assigned to the intruder, as it is assumed to have no influence on the granular gas. Therefore, the number of intruders $\mathcal{N}_0$ merely has an statistical meaning, and may therefore be chosen arbitrarily.

\begin{figure}
\begin{center}
\includegraphics[width=.7\columnwidth]{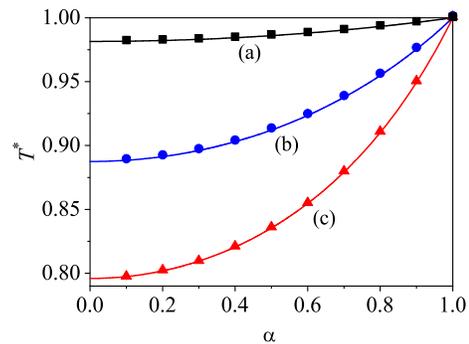}
\end{center}
\caption{Plot of the reduced granular temperature $T^*$ versus the coefficient of restitution $\al$ for a three-dimensional ($d=3$) system with $T_\text{b}^*=1$ and three different values of the solid volume fraction $\phi$: (a) $\phi=0.01$ (black line and squares); (b) $\phi=0.1$ (blue line and circles); and (c) $\phi=0.25$ (red line and triangles). The symbols refer to the DSMC results.
\label{fig2}}
\end{figure}

We measure the (reduced) granular temperatures $T^*$ and $T_0^*$ as well as the (scaled) diffusion coefficient $D^*$ in the homogeneous steady state.
The temperatures are computed from the masses and velocities, whereas $D^*$ is obtained from the ensemble-averaged square deviation
of the intruder's position [Eq.\ \eqref{4.29}]:
\beq
\label{5.1}
D=\frac{1}{2 d \Delta t}\Big[\langle|\mathbf{r}_0(t+\Delta t)-\mathbf{r}_0(0)|^2\rangle-\langle|\mathbf{r}_0(t)-\mathbf{r}_0(0)|^2\rangle,
\eeq
where $|\mathbf{r}_i(t)-\mathbf{r}_i(0)|$ is the distance traveled by the intruder up to time $t$. Here, $\langle \cdots \rangle$
denotes the average over the $\mathcal{N}_0$ intruders and $\Delta t$ is the time step. In our simulations, we have followed a procedure similar
to that used by Montanero and Garz\'o \cite{MG02} (who performed simulations for freely cooling granular mixtures) to numerically
solve the Enskog--Lorentz kinetic equation under steady conditions.
We have simulated a system constituted by a total number of $\mathcal{N} = 10^5$ inelastic, smooth, hard spheres, of which $\mathcal{N}_0=4\times 10^4$
are tracer (or intruder) particles. For sufficiently rarefied gases, the collisions are assumed to be instantaneous, and so the free flight of particles
decouples in time from the collision stage. The DSMC method maintains this assumption and can therefore be divided into two steps:
the convective and the collision stages. The latter refers to the interparticle collisions, whereas in the convective stage particles of each
component change their velocities due to the interactions with the bath. For a three-dimensional
system ($d=3$), the influence of the interstitial fluid on grains is taken into account by updating the velocity $\mathbf{v}_k$ of every
single grain of each species $i$ after each time step $\Delta t$ according to the rule \cite{KG14}:
\beq
\label{DSMC}
\mathbf{v}_k\to e^{-\gamma_i\Delta t}\mathbf{v}_k+\left(\frac{6\gamma_i T_\text{b} \Delta t}{m_i}\right)^{1/2}\textbf{R}_k.
\eeq
Here, $\textbf{R}_k$ is a random vector of zero mean and unit variance. Equation \eqref{DSMC} converges to the Fokker--Plank operator
when the time step $\Delta t$ is much shorter than the mean free time between collisions \cite{KG14}. The procedure is replicated twenty times, whereby each of replicas comprises up to $10^3$ intruder-grain collisions that are counted once the steady regime has been reached.

\subsection{Self-diffusion case}

\begin{figure}
\begin{center}
\includegraphics[width=.7\columnwidth]{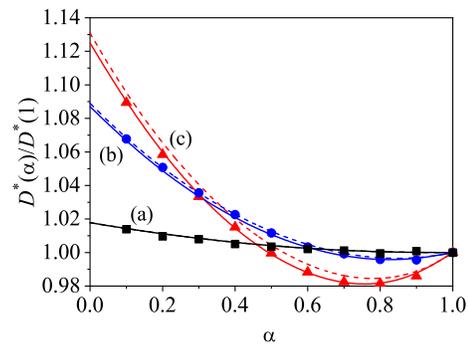}
\end{center}
\caption{Plot of the (reduced) self-diffusion coefficient $D^*(\al)/D^*(1)$ versus the coefficient of restitution $\al$ for a three-dimensional ($d=3$) system with $T_\text{b}^*=1$ and three different values of the solid volume fraction $\phi$: (a) $\phi=0.01$ (black lines and squares); (b) $\phi=0.1$ (blue lines and circles); and (c) $\phi=0.25$ (red lines and triangles). The symbols refer to the DSMC results while the solid (dashed) lines correspond to the theoretical results obtained from the second (first) Sonine approximation. Here, $D^*(1)$ denotes the elastic-limit value of the tracer self-diffusion coefficient consistently obtained in each approximation.
\label{fig3}}
\end{figure}

Let us first consider the self-diffusion case. In this limit case, $T^*=T_0^*$ for arbitrary values of $\al$. For the case of hard spheres, a good approximation to $\chi$ is \cite{CS69}
\beq
\label{5.2}
\chi(\phi)=\frac{1-\frac{1}{2}\phi}{(1-\phi)^3}.
\eeq
Moreover, for the sake of illustration, we consider the following
expression for $R(\phi)$ obtained from simulations for hard spheres systems  \cite{HBK05,BHK07,YS09b}:
\beq
\label{5.3}
R(\phi)=\frac{10 \phi}{(1-\phi)}+\left(1-\phi\right)^3\left(1+1.5 \sqrt{\phi}\right).
\eeq
With the specifications given by Eq.\ \eqref{5.2} and Eq.\ \eqref{5.3}, we are now in the position to compute the reduced temperature $T^*=x_\text{phys}^2\, T_\text{b}^*$ via the cubic equation \eqref{2.12}.
Figure \ref{fig2} depicts $T^*$
versus the coefficient of restitution $\al$ for a three-dimensional ($d=3$) system with $T_\text{b}^*=1$ and three different values of the solid volume fraction $\phi$. The curves correspond to the solution provided by Eq.\ \eqref{2.12}, while the symbols represent our DSMC results.
The dependence of  $T^*$ on both $\al$ and $\phi$ is as expected. Thus, for a given density $\phi$, when $\alpha$ grows, the energy
dissipated in the collisions decreases, and so the kinetic energy of grains (or equivalently, their reduced temperature $T^*$) increases.
Furthermore, for a given value of $\al$, an increase in density leads to an increase in the collision frequency, and thus to a decrease in
the  mean kinetic energy of grains (implying a lower temperature). Figure \ref{fig2} also highlights the excellent agreement between theory
and simulations, even for strong inelasticity (small $\al$) and/or large density values.

\begin{figure}
\begin{center}
\includegraphics[width=.7\columnwidth]{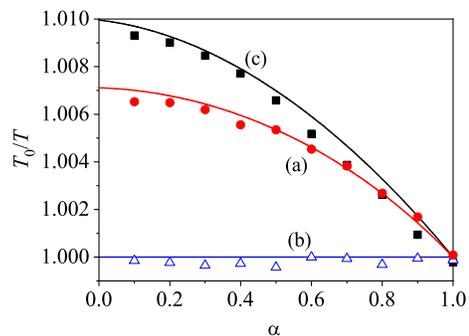}
\end{center}
\caption{Plot of the temperature ratio $T_0/T$  versus the (common) coefficient of restitution $\al$ for a three-dimensional ($d=3$) system with $T_\text{b}^*=1$, $\phi=0.1$, and three different mixtures: (a) $m_0/m=0.5$ $\sigma_0/\sigma=(0.5)^{1/3}$ (red line and circles); (b) $m_0/m=1$ and $\sigma_0/\sigma=1$ (blue line and triangles); and (c) $m_0/m=8$ and $\sigma_0/\sigma=2$ (black line and squares). The symbols refer to the DSMC results.
\label{fig4}}
\end{figure}

\begin{figure}
\begin{center}
\includegraphics[width=.7\columnwidth]{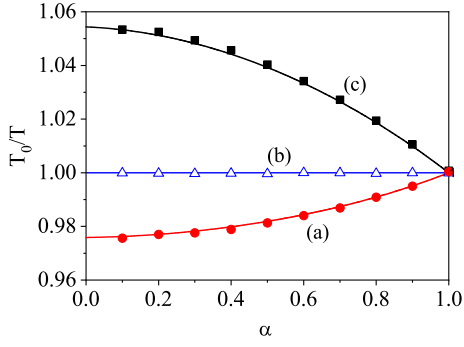}
\end{center}
\caption{Plot of the temperature ratio $T_0/T$ versus the (common) coefficient of restitution $\al$ for a three-dimensional ($d=3$) system with $T_\text{b}^*=1$, $\phi=0.1$, and three different mixtures: (a) $m_0/m=0.5$ and $\sigma_0/\sigma=0.5$ (red line and circles); (b) $m_0/m=1$ and $\sigma_0/\sigma=1$ (blue line and triangles); and (c) $m_0/m=10$ and $\sigma_0/\sigma=5$ (black line and squares). The symbols refer to the DSMC results.
\label{fig5}}
\end{figure}

The $\al$-dependence of the self-diffusion coefficient $D^*(\al)$ scaled with respect to its value $D^*(1)$ in the elastic limit
is plotted in Fig.\ \ref{fig3} for the same systems as in Fig.\ \ref{fig2}. Theoretical predictions provided by the first and second Sonine approximations are compared against DSMC simulations based on the Einstein form \eqref{5.1}. For very dilute gases ($\phi=0.01$),
the ratio $D^*(\al)/D^*(1)$ decreases monotonically with increasing coefficient of restitution $\al$,
but for  moderate densities a non-monotonic dependence on $\al$ is seen to emerge.  

While the first-Sonine solution compares quite well with simulations in the low-density regime, at high densities ($\phi=0.25$) small discrepancies appear. These differences are clearly mitigated by
the second-Sonine solution; its prediction yields an excellent agreement with the DSMC results, even for rather strong inelasticity.

Replacing $\sqrt{T^*}$ with $x_\text{phys} \sqrt{T_\text{b}^*}$ in Eq.\ \eqref{4.22} and expanding the resulting expression in powers of $(1-\alpha)$, it is possible to obtain an estimate for the ratio $D^*(\alpha)/D^*(1)$ calculated in the first Sonine approximation for the quasielastic regime ($\al \lesssim 1$). We omit the details of the rather tedious calculation and give directly the result to quadratic order in $(1-\alpha)$: 
\beq
\label{diff-ratio}
\frac{D^*(\alpha)}{D^*(1)}\!\approx\! 1-\frac{4A^2}{1\!+\!4A}(1-\alpha)+
\frac{A\!+\!10A^2\!+\!30A^3\!+\!40A^4}{(1\!+\!4A)^2}(1-\alpha)^2 
\eeq
with $A=\sqrt{T_\text{b}^*}/(\lambda d)$. Requiring that the $\alpha$-derivative
of Eq.\ \eqref{diff-ratio} vanishes, we obtain an estimate of the $\alpha$-value 
for which a minimum of the self-diffusion coefficient is attained (cf.\ dashed lines in Fig.\ \ref{fig3}):
\beq
\label{alpha-min}
\alpha_\text{min}\approx 1-\frac{2A+8A^2}{1+10A+30A^2+40A^3}
\eeq
The above estimate lies close to the actual value of $\alpha_\text{min}$ as long as one remains in the quasielastic regime. In Sec. \ref{sec6}, we will revisit Eq.\ \eqref{alpha-min} when we discuss the physics underlying the non-monotonic behavior of the diffusion coefficient.

\subsection{Diffusion case}

Consider now a setting in which intruder and grains are mechanically different (they may differ in size and mass as well as in their
coefficients of restitution). To reduce the size of the parameter space, we consider a three-dimensional system in which both intruders and
grains have a common coefficient of normal restitution $\al=\al_0$. For $d=3$, a good approximation for $\chi_0$ is \cite{GH72}
\beq
\label{5.4}
\chi_0=\frac{1}{1-\phi}+3 \frac{\sigma_0}{\sigma+\sigma_0}\frac{\phi}{(1-\phi)^2}+2 \left(\frac{\sigma_0}{\sigma+\sigma_0}\right)^2 \frac{\phi^2}{(1-\phi)^3}.
\eeq
In addition, for an interstitial fluid with low-Reynolds-number and moderate densities, computer simulations for polydisperse gas-solid flows
provide a reasonable estimate for $R_0$, namely, \cite{HBK05,BHK07,YS09b}
\beq
\label{5.5}
R_0=1+\left(R-1\right)\left[a \frac{\sigma_0}{\sigma}+(1-a)\frac{\sigma_0^2}{\sigma^2}\right],
\eeq
where
\beq
\label{5.6}
a(\phi)=1-2.660\phi+9.096\phi^2-11.338\phi^3.
\eeq
Note that for mechanically equivalent particles, one has $R_0=R$, as it should be the case.

\begin{figure}
\begin{center}
\includegraphics[width=.7\columnwidth]{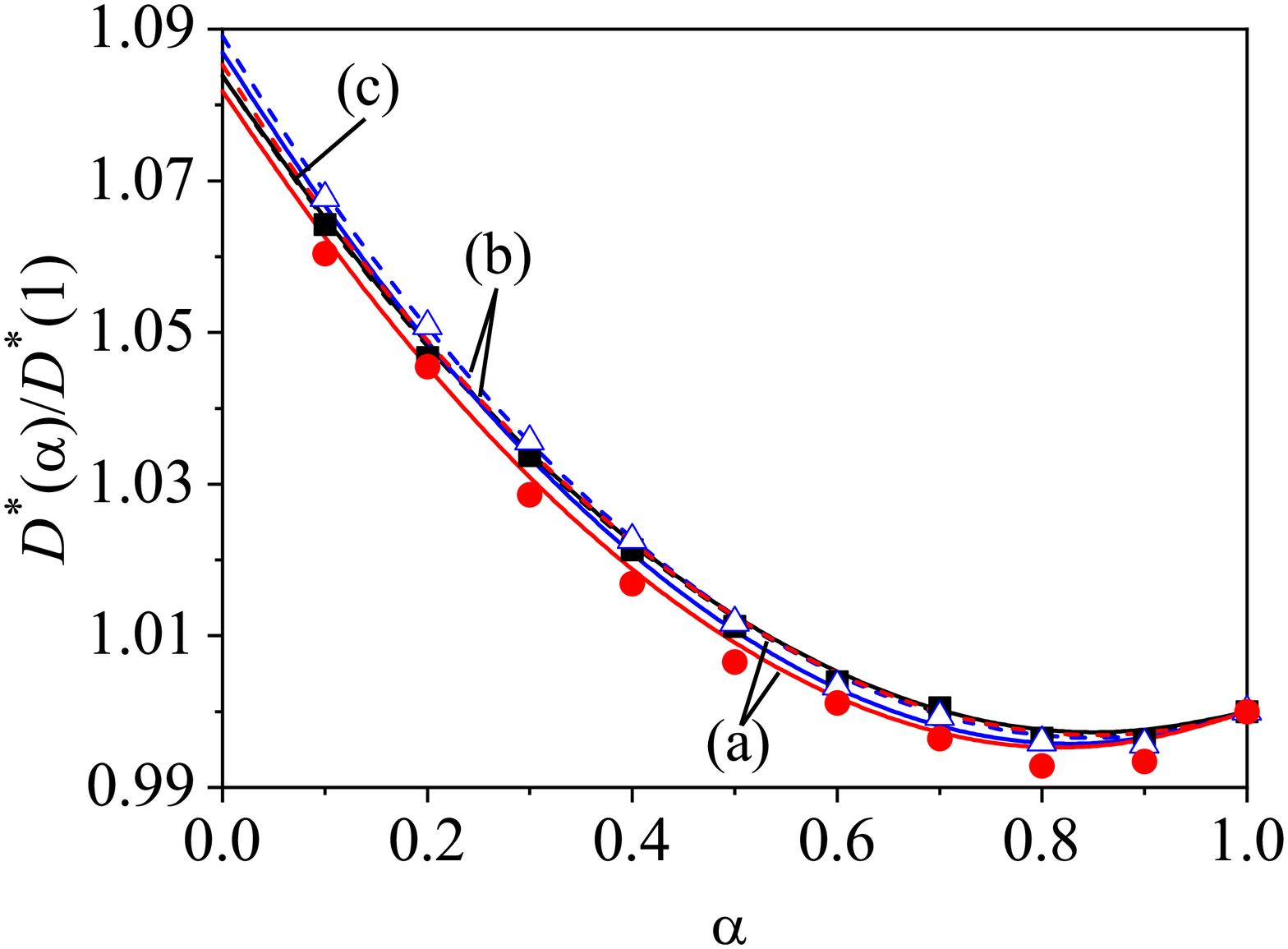}
\end{center}
\caption{Plot of the (reduced) diffusion coefficient $D^*(\al)/D^*(1)$ versus the coefficient of restitution $\al$ for a three-dimensional ($d=3$) system with $T_\text{b}^*=1$, $\phi=0.1$, and three different mixtures: (a) $m_0/m=0.5$ and $\sigma_0/\sigma=(0.5)^{1/3}$ (red lines and circles); (b) $m_0/m=1$ and $\sigma_0/\sigma=1$ (blue lines and triangles); and (c) $m_0/m=8$ and $\sigma_0/\sigma=2$ (black lines and squares). Solid and dashed lines are for the second and first Sonine approximations, respectively, while the symbols are for the DSMC results.
\label{fig6}}
\end{figure}
\begin{figure}
\begin{center}
\includegraphics[width=.7\columnwidth]{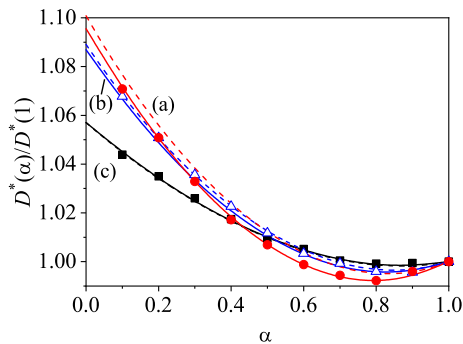}
\end{center}
\caption{Plot of the (reduced) diffusion coefficient $D^*(\al)/D^*(1)$ versus the coefficient of restitution $\al$ for a three-dimensional ($d=3$) system with $T_\text{b}^*=1$, $\phi=0.1$, and three different mixtures: (a) $m_0/m=\sigma_0/\sigma=0.5$ (red lines and circles); (b) $m_0/m=1$ and $\sigma_0/\sigma=1$ (blue lines and triangles); and (c) $m_0/m=10$ and $\sigma_0/\sigma=5$ (black lines and squares). Solid and dashed lines are for the second and first Sonine approximations, respectively, while
the symbols are for the DSMC results.
\label{fig7}}
\end{figure}

Figure \ref{fig4} shows the dependence of the temperature ratio $T_0/T$ on the common coefficient of restitution ($\al=\al_0$) for $T_\text{b}^*=1$, $\phi=0.1$, and two mixtures
consisting of particles with the same mass density [$m_0/m=(\sigma_0/\sigma)^3$]. We observe a tiny influence (amplified by the scale of
the vertical axis) of the mass and diameter ratios on $T_0/T$; in any case, the deviation
of the temperature ratio $T_0/T$ from unity is very modest; in other words, the breakdown of energy equipartition in a multicomponent
granular suspension where one of the species is present in tracer concentration does not appear to be significant. This finding contrasts with the results
derived for dry granular mixtures, where the departure of $T_0/T$ from unity becomes very important for both disparate mass or size ratios
and/or for strong inelasticity \cite{GD99b,MG02}. We also see a very good agreement between theory and simulations over the complete
range of $\al$-values.

The breakdown of energy equipartition is slightly more noticeable for the mixtures considered in Fig.\ \ref{fig5}, which do not have the same mass density. In addition, as occurs for dry
granular mixtures \cite{G19}, we observe that the temperature of the intruder is higher (lower) than that of grains when the
former is heavier (lighter) than the latter. Excellent agreement between theory and simulations is again obtained. A
broader discussion of the dependence of $T_0/T$ on the system parameters will be carried out in Sec.\ \ref{sec6} when
we analyze the impact of the temperature ratio on the effective mean free path.

Finally, let us consider the diffusion coefficient $D^*$. In Figs.\  \ref{fig6} and \ref{fig7} we plot the ratio $D^*(\al)/D^*(1)$ as a function of the (common) coefficient of restitution 
$\al$ for the systems studied in Figs.\ \ref{fig4} and \ref{fig5}, respectively. As in the case of the temperatures, we find a weak influence of the
mass and diameter ratios on the value of $D^*(\al)/D^*(1)$. In fact, in Fig.\ \ref{fig6} the first-Sonine solutions for the three chosen systems are practically indistinguishable from each other, although computer simulations do reveal small differences in the behavior. The second-Sonine solution is able to account for the observed differences between the three mixtures, and again exhibits excellent agreement with the simulations. As in the case of self-diffusion, we observe a non-monotonic dependence of $D^*$ on the coefficient of restitution $\al$. A more significant discrepancy between the first and the second Sonine approximation is observed in Fig.\ \ref{fig7} when the mass and/or diameter ratios are smaller than unity. 
 In the other case $m_0/m=10$ and $\sigma_0/\sigma=5$,  the first and second Sonine approximations practically give the same results,
 showing that the convergence of the Sonine polynomial expansion improves when $m_0/m$ and $\sigma_0/\sigma$ increase (this also
 happens in dry granular mixtures \cite{G19}).

\section{Random walk interpretation and physical discussion}
\label{sec6}

In the previous sections we have computed the MSD of intruders in granular suspensions by resorting to Enskog kinetic theory, a
rigorous, widely used method to characterize transport in molecular and granular gases \cite{CC70,G19}. A less common alternative
is the so-called free path theory \cite{CC70,J82}. In this approach,  the motion of the gas molecules is viewed as a (random) flight
between collisions.  The deflection caused by each collision is identified as a jump, and the succession of such jumps as a random
walk giving rise to a diffusive process on long enough time scales. The appeal of this approach (and the reason
why we use it below to interpret our results) lies in its ability to provide a simple, intuitive description of the diffusion process.
Specifically, our ultimate goal is to gain some physical intuition for the results we found in Sec.\ \ref{sec4} with the help of kinetic
theory.

Let $\mathbf{r}_i$ be the position of the intruder at the $i$-th collision with a grain.  We will denote by $\Bell_i$ the $i$-th displacement
between collisions: $\Bell_i=\mathbf{r}_i-\mathbf{r}_{i-1}$. Therefore, the intruder's displacement after $N$ collisions is
$\Delta \mathbf{r} =\sum_{i=1}^N \Bell_i$, and the MSD can be written as follows:
\begin{equation}
\label{MSDNele2}
\langle |\Delta \mathbf{r}|^2\rangle = N \ell_e^2.
\end{equation}
Here, we have introduced the ``effective mean free path'' $\ell_e$ (EMFP),  defined via the equation
\begin{equation}
\label{sq-EMFP}
\ell_e^2=   \langle \ell^2\rangle+ \frac{1}{N}\sum_{i\neq j}^N \langle \Bell_i\cdot\Bell_j\rangle,
\end{equation}
where $\langle \ell^2\rangle\equiv \langle \ell_1^2\rangle=\langle \ell_2^2\rangle=\ldots$ If one neglects the correlation terms
(i.e., if one takes $\langle \Bell_i\cdot\Bell_j\rangle=0$), one finds a simple yet fairly rough approximation for the MSD, namely,
\begin{equation}
\label{rw-msd}
\langle |\Delta \mathbf{r}|^2\rangle = N \langle \ell^2\rangle.
\end{equation}
For elastic hard spheres, the above expression underestimates the result obtained from Eq.\ \eqref{MSDNele2} by more than $40 \%$
(see Sec.~4 of Ref.\ \cite{ABG22}). The reason is that the jump-jump correlations $\langle \Bell_i\cdot\Bell_j\rangle$ are \emph{positive}
and add up in time to yield an important contribution to the MSD. This reflects the ``persistence'' of displacements arising from
the microscopic collision rules, which make forward collisions more likely than backward ones; the net effect being that, after collisions,
particles will tend to move forward at angles not too large with respect to their precollisional direction \cite{CC70}.  The EMFP defined above (which is larger than the actual mean free path) captures this effect, and allows one to proceed \emph{as if the steps of the random walk were isotropic} by using the expression \ \eqref{MSDNele2} for the MSD.

The exact microscopic evaluation of the correlations $\langle \Bell_i\cdot\Bell_j\rangle$ and the (squared) EMFP $\ell_e^2$ is not
an easy task, even for the simplest case of elastic hard spheres \cite{Y49}. However, we can use Eq.~\eqref{MSDNele2} and the expression \eqref{4.29} of the MSD obtained in Sec.~\ref{sec5} to estimate $\ell_e$. Note that  Eq.~\eqref{MSDNele2} can be rewritten as
\begin{equation}
\label{MSDNele2.1}
\langle |\Delta \mathbf{r}|^2(t)\rangle = s_0(t) \ell_e^2
\end{equation}
where $\langle |\Delta \mathbf{r}|^2(t)\rangle$ is the MSD up to time $t$, and $s_0(t)=\nu_0 t$  denotes the average number of intruder-grain
collisions ($\nu_0$ being the average intruder-grain collision frequency). Equivalently, when the intruder is seen as a random walker,
$s_0(t)$ represents the average number of steps taken up to time $t$. Taking into account Eqs.~\eqref{4.29} and \eqref{MSDNele2.1}, one finds
\begin{equation}
\label{x.0}
\frac{\ell_e^2}{\sigma^2}=\frac{2dD^*}{\nu_0^*},
\end{equation}
with $\nu_0^*=\nu_0/\gammaSt$. In Sec.~\ref{sec4}, the (reduced) diffusion coefficient $D^*$ has been evaluated  in the first and
second Sonine approximations. In Ref.~\cite{ABG22}, the following (approximate) expression for the (reduced) collision frequency $\nu_0^*$ was provided:
\begin{equation}
\label{nu0gaSt}
\nu_0^* = \frac{2^d d}{\sqrt{\pi}} \, \left(\frac{\overline{\sigma}}{\sigma}\right)^{d-1} \phi\,
{\chi_0}\left(\frac{1+\beta}{2\beta}\right)^{1/2}\, \sqrt{T^*},
\end{equation}
where the definition \eqref{2.6} of $\phi$ has been employed. For self-diffusion,
\beq
\label{nu0.1}
\nu_0^*=\nu^*=\frac{2^d d}{\sqrt{\pi}} \phi\chi\sqrt{T^*}.
\eeq
From Eqs.\ \eqref{x.0} and \eqref{nu0gaSt}, one eventually finds
\begin{equation}
\label{x.1}
\frac{\elee^2}{\sigma^2}=  \sqrt{\pi} \,  \left( 1+\frac{\sigma_0}{\sigma}\right)^{1-d}\,
\left(\frac{2\beta}{1+\beta}\right)^{1/2}
 \frac{D^*}{ \phi\,
{\chi_0}\, \sqrt{T^*}}.
\end{equation}

Our aim in this section is to assess the impact of the inelasticity of grains on both the diffusion coefficient $D$ and the corresponding MSD and to rationalize it with simple arguments. To this end, we will study the behavior of the ratios
\begin{equation}
\label{MSDred}
\frac{\langle |\Delta \mathbf{r}|^2(t;\al)\rangle}{\langle |\Delta \mathbf{r}|^2(t;1)\rangle}
=\frac{D^*(\alpha)}{D^*(1)}
=\frac{\nu_0(\alpha)}{\nu_0(1)}\, \frac{\ell_e^2(\alpha)}{\ell_e^2(1)}.
\end{equation}
Equation \eqref{MSDred} tell us that the change of the MSD with inelasticity can be inferred from the respective changes in the
collision frequency and in the EMFP.  To this end, we will discuss separately the case of self-diffusion and the general case with
intruders and grains differing in their mechanical properties. Henceforth, we will take $d=3$ and for the sake
of simplicity we will assume a common coefficient of restitution ($\alpha_0=\alpha$).

\subsection{Self-diffusion case}
\label{secRWSelf}

\begin{figure}
\begin{center}
\includegraphics[width=.7\columnwidth]{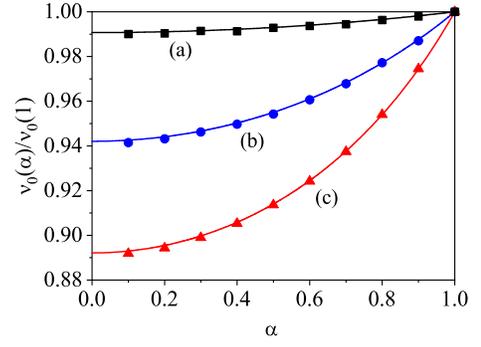}
\end{center}
\caption{Plot of the (reduced) collision intruder-grain frequency ${\nu_0(\alpha)}/{\nu_0(1)}$ versus the coefficient of restitution $\al$ for a three-dimensional ($d=3$) system with $m_0/m=\sigma_0/\sigma=1$, $T_\text{b}^*=1$,  and three  different densities: (a) $\phi=0.01$ (black line and squares); (b) $\phi=0.1$ (blue line and circles); and (c) $\phi=0.25$ (red line and triangles). The symbols are DSMC results.
\label{fignu0Auto}}
\end{figure}

In this limiting case, $\nu(\al)=\nu_0(\al)$ where 
\begin{equation}
\label{nuRedAuto}
\frac{\nu_0(\alpha)}{\nu(1)} = \sqrt{\frac{T^*(\alpha)}{T_\text{b}^*}}
\end{equation}
and  
\begin{equation}
\label{eleAuto}
\frac{\ell_e^2(\alpha)}{\ell_e^2(1)} = \frac{D^*(\alpha)/D^*(1)}{\sqrt{T^*(\alpha)/T_\text{b}^*}}.
\end{equation}
In Eqs.\ \eqref{nuRedAuto} and \eqref{eleAuto}, use has been made of Eq.\ \eqref{nu0.1} and the identity $T^*(1)=T_\text{b}^*$. According to Eq.\ \eqref{nuRedAuto}, the density dependence of the ratio
$\nu_0(\al)/\nu(1)$ is solely given by the density dependence of the (reduced) temperature $T^*$. As shown in Fig.~\ref{fignu0Auto}, 
the behavior of $\nu_0(\al)/\nu(1)$ obtained from Eq.\ \eqref{nuRedAuto} is again in excellent agreement with simulations.

\begin{figure}
\begin{center}
\includegraphics[width=.7\columnwidth]{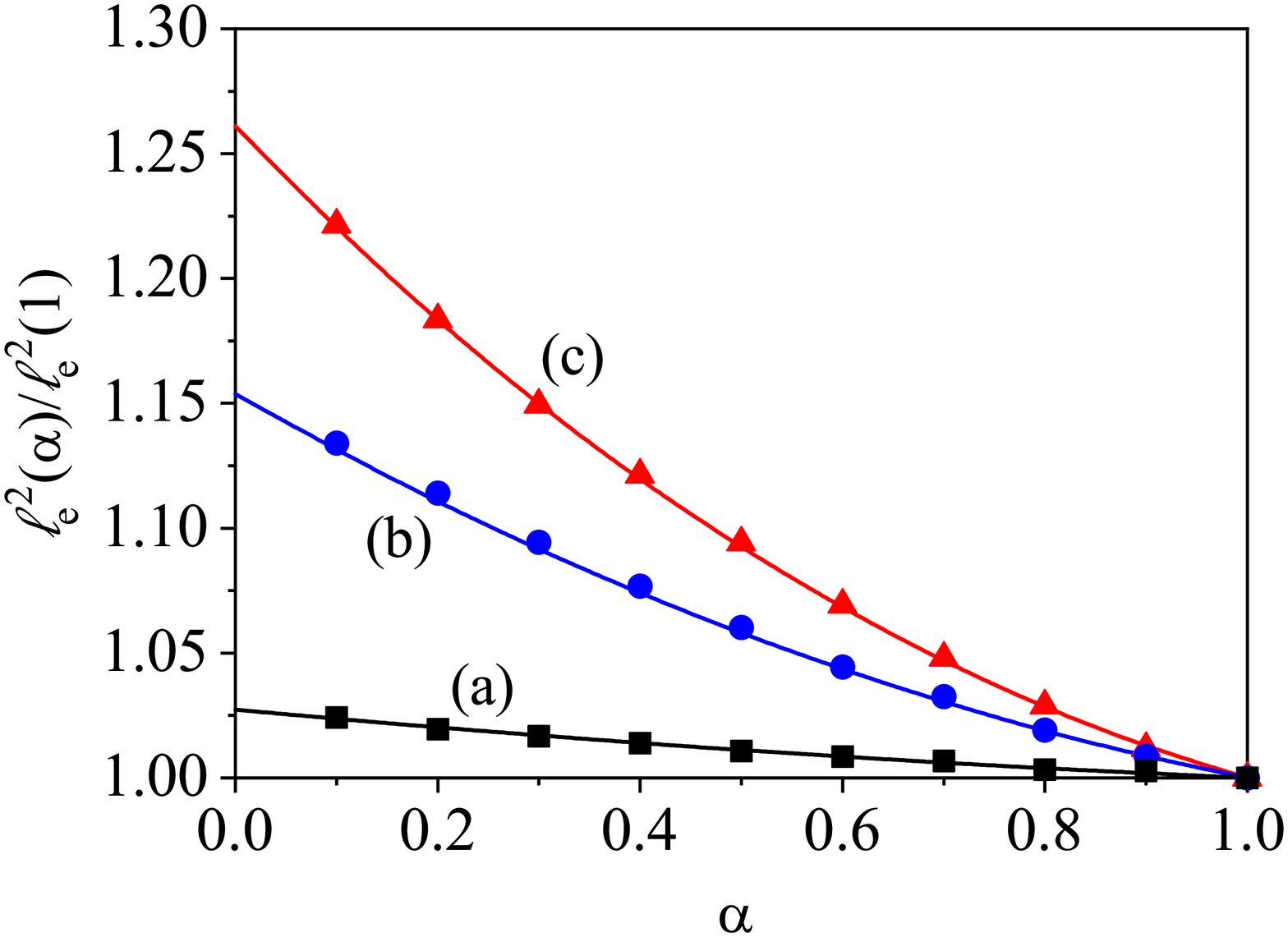}
\end{center}
\caption{Plot of  $\elee^2(\alpha)/\elee^2(1)$ versus the coefficient of restitution $\al$ for a three-dimensional ($d=3$) system with $m_0/m=\sigma_0/\sigma=1$, $T_\text{b}^*=1$,  and three  different densities: (a) $\phi=0.01$ (black line and squares); (b) $\phi=0.1$ (blue line and circles); and (c) $\phi=0.25$ (red line and triangles). The symbols are DSMC results. Theoretical results have been obtained from the second Sonine approximation.
\label{figeleAuto}}
\end{figure}

In view of Fig.~\ref{fignu0Auto}, the behavior of the ratio $D^*(\alpha)/D^*(1)$ depicted in Fig.~\ref{fig3} seems at first glance surprising,
since one could expect that the dependence of the intruder's MSD on both $\al$ and $\phi$ follows that of the collision frequency
(this is in fact what the right-hand side of Eq.\ \eqref{MSDred} tells us). For example, one might expect that the more displacements/collisions $s_0(t)=\nu_0 t$ the intruder experiences in a given time $t$, the further it will travel.
And yet we see that $\nu(\alpha)/\nu(1)$ always increases with $\alpha$ (or with the density $\phi$) for fixed
$\phi$ (or $\al$), as opposed to the behavior of the reduced diffusion coefficient $D^*(\alpha)/D^*(1)$. The explanation for this apparent contradiction lies in the behavior of the EMFP, which appears in the prefactor ${\ell_e^2(\alpha)}/{\ell_e^2(1)}$ multiplying  the
reduced collision frequency in Eq.~\eqref{MSDred}. The growth of $\ell_e^2(\alpha)$ with decreasing $\al$ shown in Fig.~\ref{figeleAuto} is explained by the aforementioned \emph{persistence} of velocities after collisions (the postcollisional
velocity of a given particle will still retain on average a significant component in the direction of its precollisional motion \cite{CC70}). Given that
collisions tend to be more focused when $\al$ becomes smaller (i.e., post-collisional velocity tends to be more parallel to the pre-collisional
velocity), the EMFP $\elee$ grows with increasing inelasticity (see Ref.\ \cite{ABG22} for more details).

There still remains to justify why the ratio ${\ell_e^2(\alpha)}/{\ell_e^2(1)}$ increases with density for fixed $\alpha$.
Increasing the density results in a larger number of collisions, to the extent that their effect on $\ell_e^2(\alpha)$
becomes more prevalent than the action exerted by the interstitial fluid on the particles. This explains why
${\ell_e^2(\alpha)}/{\ell_e^2(1)}$ increases with density for fixed $\alpha$. That is, for very low grain densities $\phi$, the influence of $\alpha$ is weaker, since the particles undergo fewer collisions per unit time.

At this stage, a comment based on the result \eqref{alpha-min}
obtained for $\alpha_\text{min}$ in the quasielastic regime is in order. Note that
the parameter $A=\sqrt{T_\text{b}^*}/(3\lambda)\propto \sqrt{T_\text{b}^*}\phi$ becomes small
for low values of the density and/or bath temperature. One then has
$\alpha_\text{min}\approx 1-2A$, i.e., for fixed $T_\text{b}^*$ the regime where $D^*(\alpha)/D^*(1)$ increases with $\alpha$ tends to vanish with decreasing density. This quantitative finding confirms the qualitative argument given 
above. On the other hand, we also see that for fixed $\phi$ a similar effect occurs as one decreases the bath temperature $T_\text{b}$, since the latter quantity is an upper bound for $T$ and the frequency of collisions goes as $\sqrt{T}$ [cf.\ Eq.\ \eqref{2.11}].

\subsection{Diffusion case}

We now consider the case where intruder and grains are mechanically different (diffusion case).  As in the self-diffusion
case, our goal here is to use Eq.~\eqref{MSDred} to gain some insight into the $\al$-dependence of the intruder's MSD.
To understand this dependence, we consider separately the behavior of the two factors ${\nu_0(\alpha)}/{\nu_0(1)}$ ${\ell_e^2(\alpha)}/{\ell_e^2(1)}$ on the rightmost part of Eq.~\eqref{MSDred}.

\subsubsection{Factor ${\nu_0(\alpha)}/{\nu_0(1)}$}
\label{sssnu}

From Eq.~\eqref{nu0gaSt}, and taking into account that $T_0(\alpha=1)=T(\alpha=1)$, one finds that $\beta=m_0/m$ and so
\beq
\label{nuFac2}
\frac{\nu_0(\alpha)}{\nu_0(1)}= \left[\frac{ T_0/T+m_0/m}{1+m_0/m}\right]^{1/2}\,\sqrt{\frac{T(\al)}{T_\text{b}}}.
\eeq
Given that the dependence of $T$ on $\alpha$ has already been studied in subsection~\ref{secRWSelf},
we will focus here on the behavior of the temperature ratio $T_0/T$ as a function of $\sigma_0/\sigma$, $m_0/m$ and $\alpha$.

At first glance, rationalizing the behavior of $T_0/T$ seems a rather difficult task, since this quantity follows as a solution of Eq.\ \eqref{3.9}, which is in fact quite involved, notably because of its dependence on the solution of Eq.\ \eqref{2.12} for T.  However, at a given temperature $T$, the behavior of the temperature ratio $T_0/T$ depends only on how the intruder temperature $T_0$ changes, which is determined solely by Eq.~\eqref{3.9}. To show the dependence of $T_0/T$ on the mass and diameter ratios in a more clear way, it is convenient to rewrite Eq.~\eqref{3.9} as
\begin{equation}
\label{Tbsimp}
\frac{T_\text{b}^*}{T_0^*}=1+ \frac{\zeta_0^*(T_0^*)}{\lambda_0} \sqrt{T^*}.
\end{equation}
This equation shows that the dependence of $T_0$ on $\sigma_0/\sigma$ and $m_0/m$ is essentially determined by the dependence of the ratio $\zeta_0^*(T_0^*)/{\lambda_0}$ on the above quantities. Although the full expression for this quantity is very cumbersome, a much simpler one can be obtained by exploiting the fact that, typically, $T_0/T\approx 1$ [see the discussion below Eq.\ \eqref{5.6} and Fig.\ \ref{fig4}]. Thus, for $d=3$, one can write
\beqa 
\label{z0la0simp}
\frac{\zeta_0^*}{\lambda_0}&\approx &\frac{2\sqrt{2}}{\sqrt{\pi}}\;\frac{\chi_0 \phi}{R_0}
\frac{\sigma }{\sigma_0 } \left(1+\frac{\sigma_0 }{\sigma }\right)^{2}
 \left(\frac{m_0}{m+ m_0}\right)^{1/2} \nonumber \\
  & &\times (1+\alpha_0) \left[1-\frac{1}{2}(1+\alpha_0)\right]
\eeqa
The behavior of $T_0$ with the diameter and mass ratios can then be easily understood from Eqs.~\eqref{Tbsimp} and \eqref{z0la0simp}.

\paragraph{Dependence of the temperature ratio $T_0/T$ on the diameter ratio $\sigma_0/\sigma$.}

\begin{figure}
\begin{center}
\includegraphics[width=.7\columnwidth]{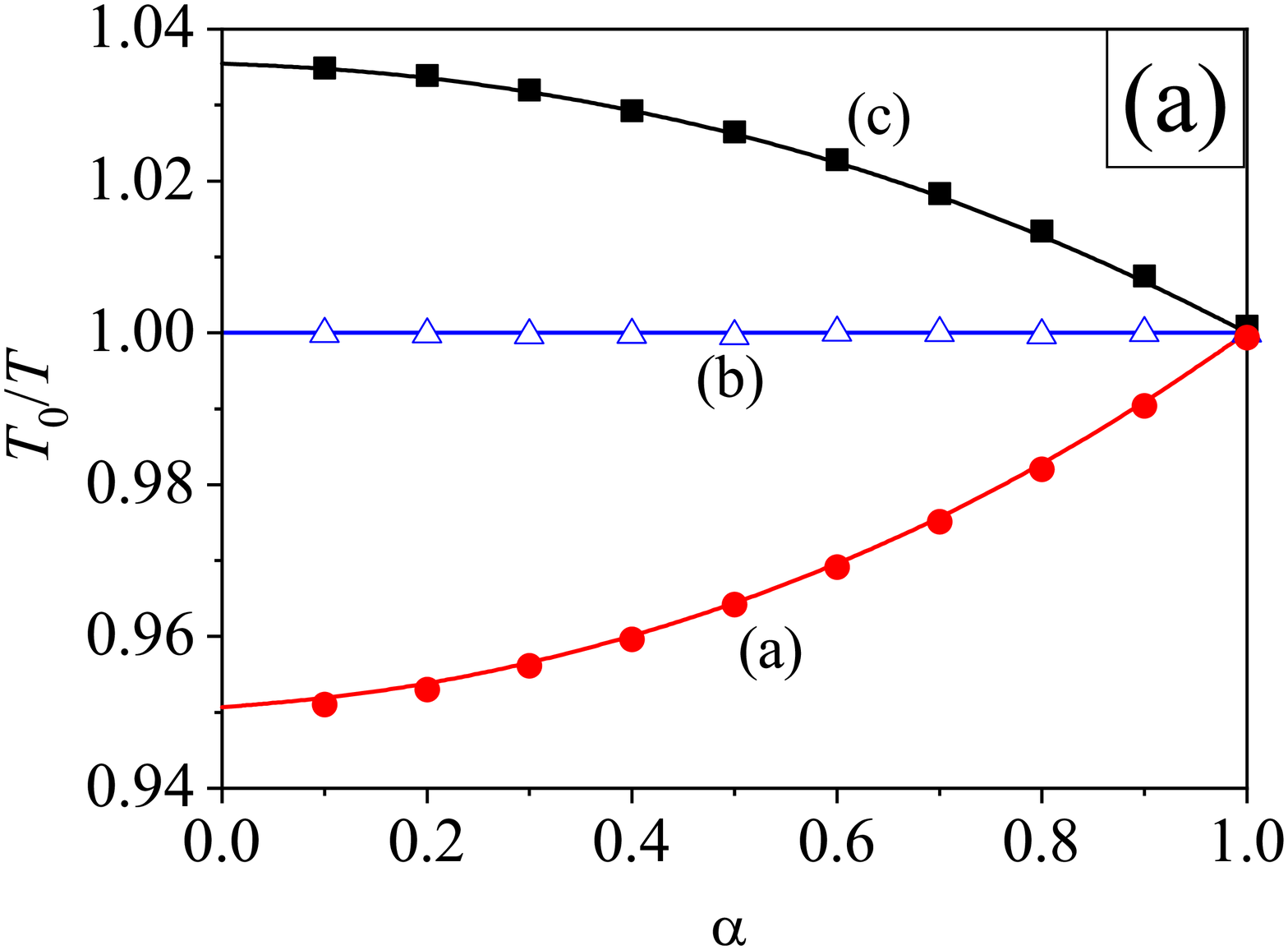} \hfill
\includegraphics[width=.7\columnwidth]{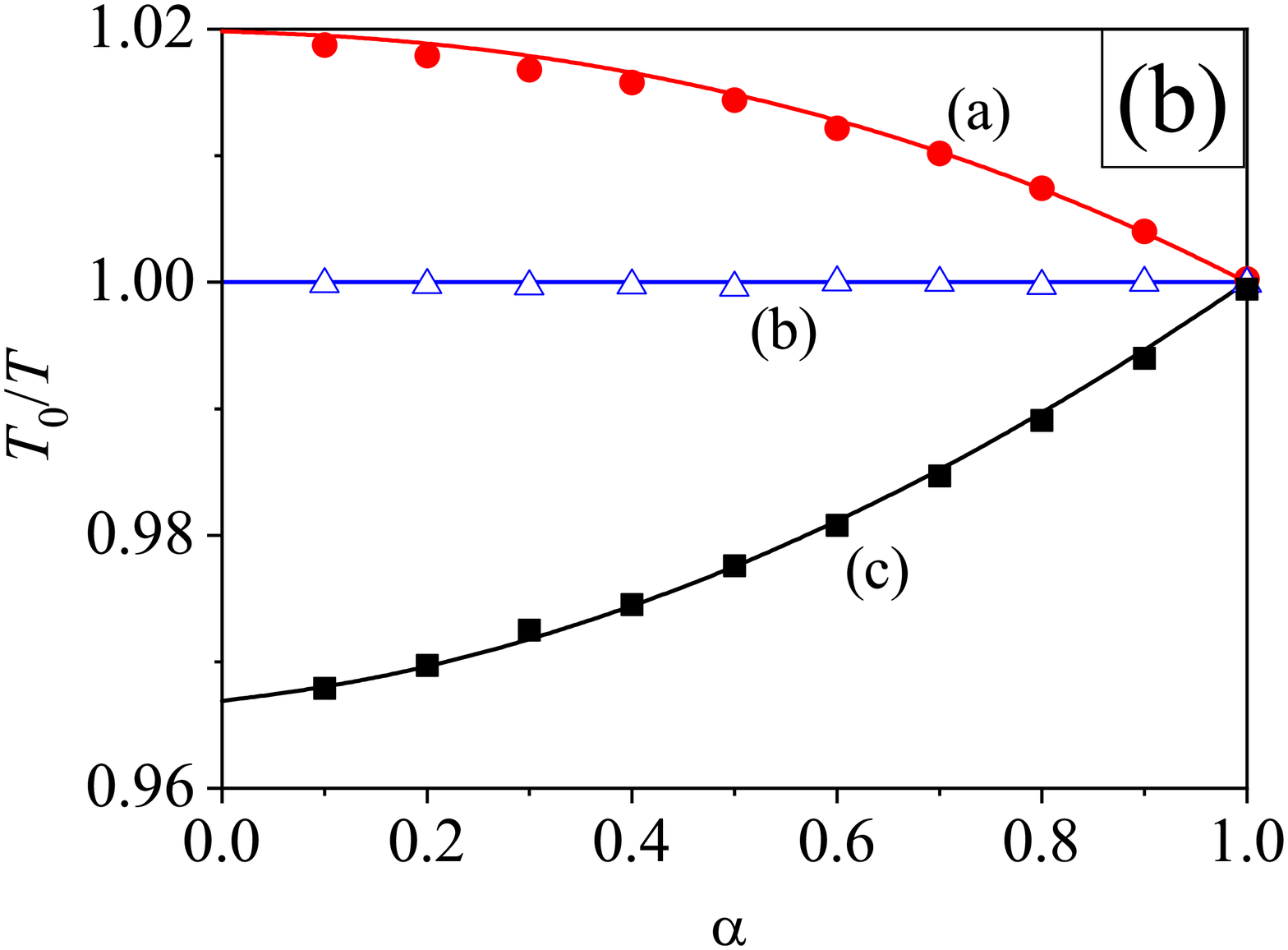}
\end{center}
\caption{Plot of the temperature ratio $T_0/T$ versus the coefficient of restitution $\al$ for $d=3$, $T_\text{b}^*=1$, $\phi=0.1$ and different mixtures: $m_0/m=1$ with   (a) $\sigma_0/\sigma=0.5$ (red line and circles); (b) $\sigma_0/\sigma=1$ (blue line and triangles); and (c) $\sigma_0/\sigma=2$ (black line and squares) [panel (a)]   and $\sigma_0/\sigma=1$ with (a) $m_0/m=0.5$ (red line and circles); (b)  $m_0/m=1$ (blue line and triangles); and (c) $m_0/m=10$ (black line and squares) [panel (b)]. The symbols are the DSMC results.  
\label{figT0T}}
\end{figure}

As for the dependence of $T_0$ on the diameter ratio $\sigma_0/\sigma$, Eqs.~\eqref{Tbsimp} and \eqref{z0la0simp}  tell us that
\begin{equation}
\label{zelawEq}
\frac{\zeta_0^*}{\lambda_0}\propto  \frac{\chi_0\left(1+\frac{\sigma_0}{\sigma}\right)^{2}}{(\sigma_0/\sigma) R_0}.
\end{equation}
Equation \eqref{zelawEq} shows that the ratio $\zeta_0^*/\lambda_0$ turns out to be a decreasing function of $\sigma_0/\sigma$ for not too small values of $\phi$ (for $\phi\gtrsim 0.015$). This explains the trends observed in the panel (a) of Fig.\ \ref{figT0T} where the temperature ratio $T_0/T$ is plotted versus $\al$ for $d=3$, $m_0/m=1$, $\phi=0.1$, and three values of $\sigma_0/\sigma$. As expected, at a given value of $\al$, $T_0/T$ increases with the diameter ratio $\sigma_0/\sigma$ since the ratio  ${\zeta_0^*}/{\lambda_0}$ is a decreasing function of $\sigma_0/\sigma$ for $\phi=0.1$.

\paragraph{Dependence of the temperature ratio $T_0/T$ on the mass ratio $m_0/m$.}

Equations.~\eqref{Tbsimp} and \eqref{z0la0simp}   tell us that
\begin{equation}
\label{zelamEq}
\frac{\zeta_0^*}{\lambda_0}\propto  \left(\frac{m_0}{m+m_0}\right)^{1/2},
\end{equation}
which is always an increasing function of $m_0/m$.  This implies that $T_0/T$ is always a decreasing function of $m_0/m$. This is confirmed in the panel (b) of Fig.\ \ref{figT0T}, which shows $T_0/T$ versus $\alpha$ for $d=3$, $\sigma_0/\sigma=1$, $\phi=0.1$, and three values of the mass ratio $m_0/m$. 
The behavior of the temperature ratio $T_0/T$ can be explained in more physical terms: the friction coefficients $\gamma$ and  $\gamma_0$ are inversely proportional to the masses of the particles 
($\gamma\sim \gamma_{\text{St}}\sim \sigma/m$ and $\gamma_0\sim \gamma_{\text{St},0}\sim \sigma_0/m_0$) and so, the effect of the bath on the temperature ratio $T_0/T$ decreases with increasing mass ratio $m_0/m$. We also note that the observed behavior in the present case of a granular suspension is markedly different from the case where the gas phase is absent (dry granular mixtures) \cite{G19}, since in the latter limiting case $T_0/T$ increases with $m_0/m$. We therefore conclude that the impact of the interstitial gas on the temperature ratio becomes very important as compared to that induced by collisions.

\paragraph{Limiting cases.}

It is instructive to estimate the factor $\nu_0(\al)/\nu_0(1)$ when the intruder is much heavier (lighter) than the grains. With respect to the temperature ratio, when $\sigma\sim \sigma_0$, we have seen that $T_0<T$ when $m_0>m$. Moreover, $T_0/T$ decreases with increasing mass ratio.
This implies that  $ (T_0/T+m_0/m)/(1+m_0/m)\approx 1$ when $m_0 \gg m$, and Eq.\ \eqref{nuFac2} yields
\begin{equation}
\label{x.2}
\frac{\nu_0(\alpha)}{\nu_0(1)}\to  \sqrt{ \frac{T(\al)}{T_\text{b}} }.
\end{equation}
Thus, in this regime ($m_0/m\gg 1$), we find that the ratio ${\nu_0(\alpha)}/{\nu_0(1)}$ essentially depends on the temperature of the grains only.
A very massive intruder will move very slowly (it will be practically at rest); therefore, the frequency of intruder-grain collisions will
essentially depend only on how fast the grains move (i.e., on the granular temperature $T$). An analogous argument applies in the limit of a very light
intruder:  $T_0/T$ is larger than 1 for $m_0<m$ and increases with decreasing $m_0/m$. Thus,
$ (T_0/T+m_0/m)/(1+m_0/m) \to T_0/T$ when $m\gg m_0$. In this limiting case, Eq.\ \eqref{nuFac2} leads to
\begin{equation}
\label{x.3}
\frac{\nu_0(\alpha)}{\nu_0(1)}  \to  \sqrt{ \frac{T_0}{T_\text{b}}}
\end{equation}
i.e., ${\nu_0(\alpha)}/{\nu_0(1)}$ depends mainly on the intruder's temperature. This makes sense: as we found previously, the breakdown
of energy equipartition is weak ($T_0\sim T$) in this case, implying that a very light intruder must move very fast in comparison with the
grains to ensure that the ratio $T_0/T$ does not deviate much from 1. In turn, this means that the frequency of intruder-grain
collisions will essentially depend only on how fast the intruder moves (i.e., on the intruder temperature $T_0$).

\paragraph{Temperature and collision frequency in mingled cases.}

A simultaneous increase or decrease in diameter and mass ratios gives rise to competing effects at the level of $T_0/T$ and $\nu_0(\al)/\nu_0(1)$. Thus, it is in general difficult to predict which of the two effects is the dominant one.
This is illustrated in Figs.~\ref{fig4} and \ref{fig5}.
In particular, taking as reference the self-diffusion case, we see in Fig.~\ref{fig4} (for the case $\{\sigma_0/\sigma=0.5^{1/3}\approx 0.79,
m_0/m=0.5\}$) that a small reduction in the diameter ratio (by a factor of $\approx 0. 79$) is not able to counterbalance a large
reduction in the mass ratio (by a factor of 0.5); consequently, the curve corresponding to this case lies above the self-diffusion curve ($\sigma_0/\sigma=1,m_0/m=1$), but below the curve for $\{\sigma_0/\sigma=2, m_0/m=8\}$. In contrast, Fig.~\ref{fig5} shows
that the change in the diameter ratio $\sigma_0/\sigma$ dominates over the corresponding changes in $m_0/m$; as a result of this,
larger values of $\sigma_0/\sigma$ lead to larger $T_0/T$.     The difference in scale between the cases of Figs.~\ref{fig4} and \ref{fig5}
is remarkable: in Fig.~\ref{fig4}, the intruder mass density is constant, and the departure of $T_0/T$ from unity remains below 1\%.

\begin{figure}
\begin{center}
\includegraphics[width=.7\columnwidth]{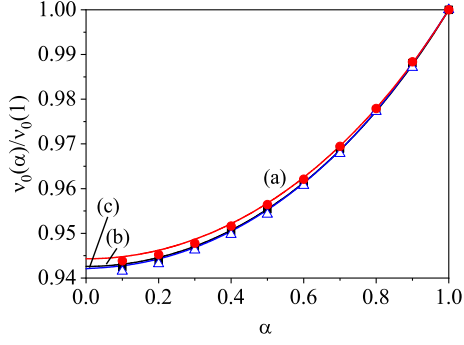}
\end{center}
\caption{\label{fign0nmix4}
Plot of $\nu_0(\alpha)/\nu_0(1)$ versus the coefficient of restitution $\al$ for $d=3$, $T_\text{b}^*=1$, $\phi=0.1$ and three different mixtures: (a) $m_0/m=0.5$ and $\sigma_0/\sigma=0.5^{1/3}$ (red line and circles); (b) $m_0/m=\sigma_0/\sigma=1$ (blue line and triangles); and (c) $m_0/m=8$ and $\sigma_0/\sigma=2$ (black line and squares). The symbols are the DSMC results.
}
\end{figure}

\begin{figure}
\begin{center}
\includegraphics[width=.7\columnwidth]{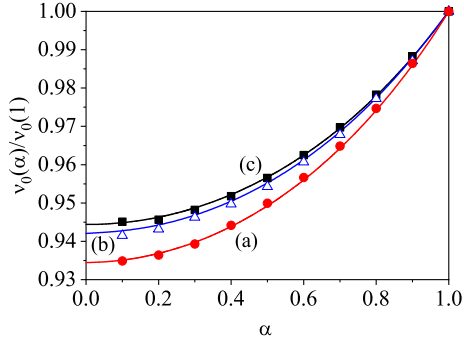}
\end{center}
\caption{\label{fign0nmix5}
   Plot of $\nu_0(\alpha)/\nu_0(1)$ versus the coefficient of restitution $\al$ for $d=3$, $T_\text{b}^*=1$, $\phi=0.1$ and three different mixtures:
 (a)  $m_0/m=\sigma_0/\sigma=0.5$ (red line and circles); (b) $m_0/m=\sigma_0/\sigma=1$ (blue line and triangles); and  (c) $m_0/m=10$ and $\sigma_0/\sigma=5$ (black line and squares).  The symbols are the DSMC results. 
}
\end{figure}

Taken together with Eq.\ \eqref{nuFac2}, the results depicted in Figs.~\ref{fig4} and \ref{fig5}  fully explain the behavior of
$\nu_0(\alpha)/\nu_0(1)$ illustrated in Figs.~\ref{fign0nmix4} and \ref{fign0nmix5} for the same systems.  For example, since the two
curves depicted in Fig.~\ref{fig4} for the temperature ratio $T_0/T$ lie close to each other, so do the corresponding curves
shown in Fig.~\ref{fign0nmix4} for $\nu_0(\alpha)/\nu_0(1)$ too. 
In contrast, despite the large separation observed in Fig.~\ref{fig5}
between the $T_0/T$-curves for $\{\sigma_0/\sigma=0.5, m_0/m=0.5\}$ $\{\sigma_0/\sigma=1,m_0/m=1\}$  and  $\{\sigma_0/\sigma=5,
m_0/m=10\}$  the corresponding curves for $\nu_0(\alpha)/\nu_0(1)$  are seen to lie close to each other (cf.\ Fig.~\ref{fign0nmix5}).
 Equation \eqref{nuFac2} provides the explanation for this behavior, as it tells us that large differences in $T_0/T$
 are strongly reduced at the level of $\nu_0(\alpha)/\nu_0(1)$ for large values of the mass ratio (which is the case here, since $m_0/m=10$).


\subsubsection{Factor $\ell_e^2(\alpha)/\ell_e^2(1)$}
\label{sssell}

\begin{figure}[t]
\begin{center}
        \includegraphics[width=0.7\columnwidth,angle=0]{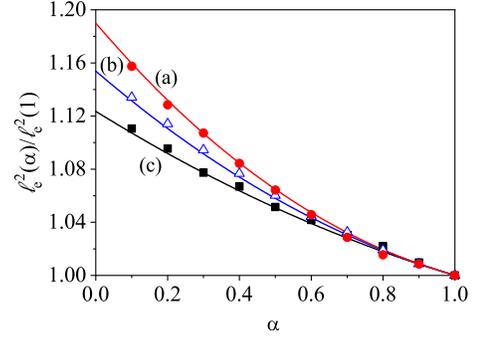}
\end{center}
\caption{\label{figele1s}
Plot of  the ratio $\ell_e^2(\alpha)/\ell_e^2(1)$ versus the coefficient of restitution $\al$ for $d=3$, $T_\text{b}^*=1$, $\phi=0.1$, $m_0/m=1$, and three different values of the diameter ratio: (a) $\sigma_0/\sigma=0.5$ (red line and circles); (b) $\sigma_0/\sigma=1$ (blue line and triangles); and  (c) $\sigma_0/\sigma=2$ (black line and squares). The symbols are the DSMC results. Theoretical results have been obtained from the second Sonine approximation.  
}
\end{figure}

\begin{figure}[t]
\begin{center}
         \includegraphics[width=0.7\columnwidth,angle=0]{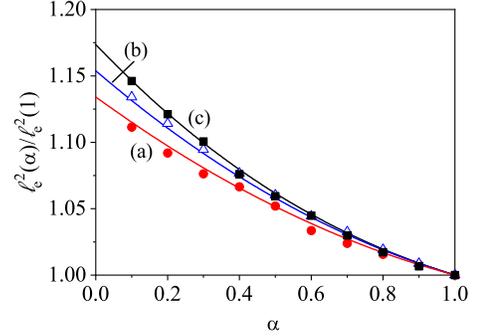}
\end{center}
\caption{\label{figele1m}
Plot of  the ratio $\ell_e^2(\alpha)/\ell_e^2(1)$ versus the coefficient of restitution $\al$ for $d=3$, $T_\text{b}^*=1$, $\phi=0.1$, $\sigma_0/\sigma=1$, and three different values of the mass ratio: (a) $m_0/m=0.5$ (red line and circles); (b) $m_0/m=1$ (blue line and triangles); and (c) $m_0/m=10$ (black line and squares). The symbols are the DSMC results. Theoretical results have been obtained from the second Sonine approximation.
}
\end{figure}

As seen in Figs.~\ref{figele1s} and \ref{figele1m}, the ratio $\ell_e^2(\alpha)/\ell_e^2(1)$ (and therefore the persistence of collisions)
decreases with increasing $\al$ \cite{BP04,G19}. We also see in Fig.~\ref{figele1s} that for fixed $\al$ the ratio $\ell_e^2(\alpha)/\ell_e^2(1)$ approaches to 1 when $\sigma_0/\sigma$ increases.  This can be ascribed to the corresponding growth of the friction coefficient
$\gamma_0\sim \sigma_0/m_0$, which signals that the influence of the interstitial gas on $\ell_e^2$ becomes increasingly relevant in
comparison with collisional effects. 
On the other hand, we see in Fig.~\ref{figele1m} that $\ell_e^2(\alpha)/\ell_e^2(1)$
grows with the mass ratio $m_0/m$ at fixed $\alpha$, as opposed to the decrease observed when $\sigma_0/\sigma$ is increased. This is
the result one intuitively expects, since the intruder's motion becomes more persistent as it gets heavier.


\begin{figure}
\begin{center}
        \includegraphics[width=0.7\columnwidth,angle=0]{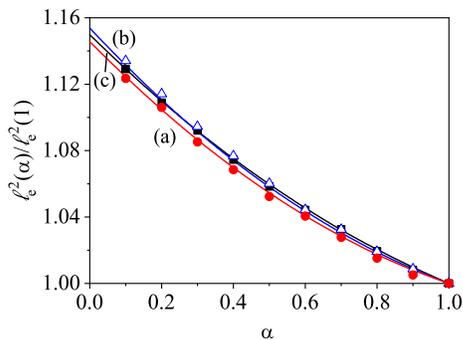}
\end{center}
\caption{\label{figele1mix4}
Plot of the ratio $\ell_e^2(\alpha)/\ell_e^2(1)$ versus the coefficient of restitution $\al$ for $d=3$, $T_\text{b}^*=1$, $\phi=0.1$ and three different mixtures: (a) $m_0/m=0.5$ and $\sigma_0/\sigma=0.5^{1/3}$ (red line and circles); (b) $m_0/m=\sigma_0/\sigma=1$ (blue line and triangles); and (c)  $m_0/m=8$ and $\sigma_0/\sigma=2$ (black line and squares). The symbols are the DSMC results. Theoretical results have been obtained from the second Sonine approximation.
}
\end{figure}

\begin{figure}
\begin{center}
        \includegraphics[width=0.7\columnwidth,angle=0]{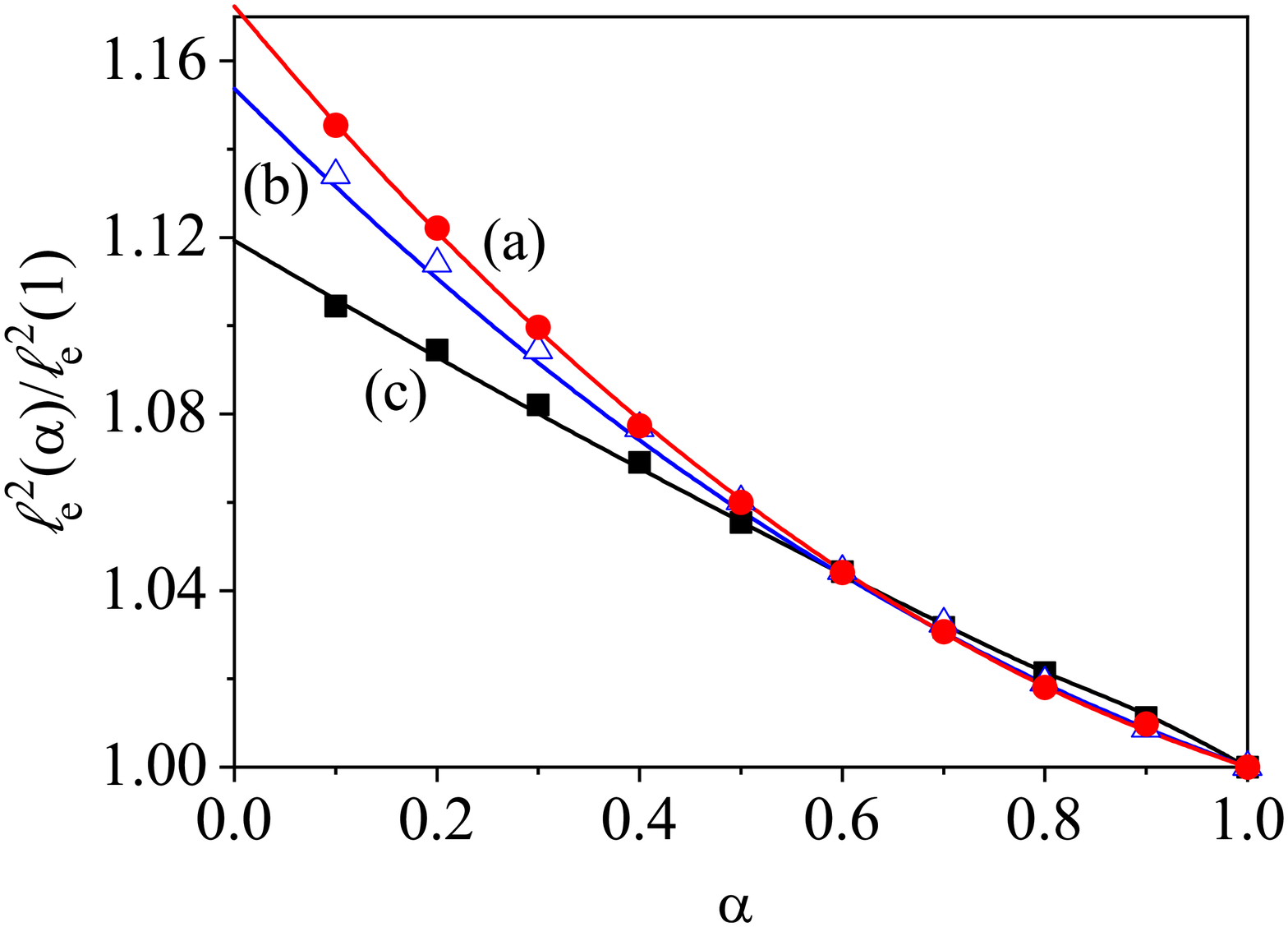}
\end{center}
\caption{\label{figele1mix5}
     Plot of the ratio $\ell_e^2(\alpha)/\ell_e^2(1)$ versus the coefficient of restitution $\al$ for $d=3$, $T_\text{b}^*=1$, $\phi=0.1$ and three different mixtures:  (a) $m_0/m=\sigma_0/\sigma=0.5$ (red line and circles); (b) $m_0/m=\sigma_0/\sigma=1$ (blue line and triangles); and (c) $m_0/m=10$ and $\sigma_0/\sigma=5$ (black line and squares). The symbols are the DSMC results. Theoretical results have been obtained from the second Sonine approximation.
}
\end{figure}

As mentioned before, when the diameter and the mass ratios are increased or decreased at the same time, the net effect of such simultaneous changes on the ratio $\ell_e^2(\alpha)/\ell_e^2(1)$ is generally difficult to predict, since they act in opposite directions.  This is illustrated
in Figs.\ \ref{figele1mix4} and \ref{figele1mix5}, where we consider the mixed cases of Figs.~\ref{fig4} and \ref{fig5}. As Fig.~\ref{figele1mix4} shows, changes in mass and size counteract each other in such a way that $\ell_e^2(\alpha)/\ell_e^2(1)$ hardly changes
(recall that the mass density is kept constant here). In contrast, the curves depicted in Fig.~\ref{figele1mix5} correspond to cases in which
an increase in the intruder's mass does not fully offset the increase in its diameter. In particular, according to the results displayed in Fig.~\ref{figele1m}, one would expect the curve for $\sigma_0/\sigma=5$ in Fig.~\ref{figele1mix5} to be much more distant from the
self-diffusion curve (dashed curve); this is not the case because the downward ``thrust'' that tends to separate the $\sigma_0/\sigma=5$
curve from the self-diffusion curve is partially offset by the upward ``thrust'' of $m_0/m=10$.


\subsection{Reduced diffusion coefficient $D(\alpha)/D(1)$ and MSD}

The results derived in subsections \ref{sssnu} and  \ref{sssell} for the ratios $\nu_0(\alpha)/\nu_0(1)$   and  $\ell_e^2(\alpha)/\ell_e^2(1)$
along with Eq.~\eqref{MSDred} fully explain the $\al$-dependence of the (reduced) diffusion coefficient $D(\alpha)/D(1)$ (or, equivalently, 
of the reduced MSD $\langle |\Delta \mathbf{r}|^2(t;\al)\rangle/\langle |\Delta \mathbf{r}|^2(t;1)\rangle$) displayed in 
Figs.~\ref{fig6} and \ref{fig7}. To understand this dependence, one should take into account that, by virtue of Eq.~\eqref{MSDred}, the
respective behaviors of the reduced collision frequency and the reduced square EMFP shown in Figs.~\ref{fign0nmix4} and  \ref{figele1mix4} 
determine the $\alpha$-dependence of $D(\alpha)/D(1)$ illustrated in Fig.~\ref{fig6}.  Similarly, Figs.~\ref{fign0nmix5} and \ref{figele1mix5} determine 
the results shown in Fig.~\ref{fig7}. For instance, the proximity of the curves corresponding to
the ratios $\nu_0(\alpha)/\nu_0(1)$ and $ \ell_e^2(\alpha)/\ell(1)$ in Figs.~\ref{fign0nmix4} and \ref{figele1mix4}
explains the proximity of the curves plotted in Fig.~\ref{fig6} for the ratio $D(\alpha)/D(1)$ .

Similarly, in Fig.~\ref{figele1mix5} we see that the relatively slow decay of $\ell_e^2(\alpha)/\ell_e^2(1)$  with increasing $\al$ 
for quasielastic systems ($\alpha\sim 1$) is outweighted by the rapid growth of $\nu_0(\alpha)/\nu_0(1)$ in this region.
This explains the (slow) growth of $D(\alpha)/D(1)$ in this quasielastic regime. On the other hand, for extremely
large  inelasticities (very small values of $\alpha$), the ratio $\nu_0(\alpha)/\nu_0(1)$ exhibits a very weak dependence
on $\al$ (the curves are nearly horizontal), and so the decrease of $\ell_e^2(\alpha)/\ell_e^2(1)$ is the dominant
effect. As a consequence, $D(\alpha)/D(1)$ decreases with increasing $\al$ in the high inelasticity region. Thus,
the non-monotonicity of the (reduced) diffusion coefficient $D(\alpha)/D(1)$ can be explained by the competition
between the decreasing function $\ell_e^2(\alpha)/\ell_e^2(1)$ and the increasing function $\nu_0(\alpha)/\nu_0(1)$.

\section{Applicability of the suspension model to real systems}
\label{secAppli}

We recall that the results obtained in this work involve different assumptions and approximations. To start with, we have restricted ourselves to low-Reynolds numbers. In addition, we have considered the effect of the force exerted by the interstitial fluid comparable to that of collisions  (as measured by the Stokes number). In this context, the question then is to what extent systems subject to the above two restrictions can both be found in nature and replicated in the laboratory. In the remainder of this section, we will attempt to provide an answer within a simplified framework that invokes several dimensionless numbers for monodisperse granular suspensions: the Reynolds $\text{Re}$ and the Stokes $\text{St}$ numbers, the (dimensionless) Stokes friction coefficient $\gamma_{\text{St}}/\nu\equiv \gamma^*/R$ as well as the reduced temperatures $\mathcal{T}$ and $T_\text{b}^*$. We will consider values of the above dimensionless quantities  for several realistic suspensions. As we will see, the resulting values fall within the ranges considered in previous sections of this paper for the pertinent quantities in our model.

The Reynolds number is  defined as
\cite{RDS16}
\beq\label{VII1}
\text{Re}=\frac{\rho_g(1-\phi)\sigma\Delta\mathbf{U}}{\eta_g},
\eeq
where $\rho_g$ is the fluid density, $\eta_g$ is the dynamic fluid viscosity, and $\Delta\mathbf{U}$ is the slip velocity, defined as the difference between the fluid velocity and the particle velocity. Here, we  assume that $\Delta\mathbf{U}$ is of the order of the thermal velocity ($\Delta\mathbf{U}\sim\sqrt{2T/m}$). The Reynolds number represent the ratio between the inertial forces and the viscous forces; it can be used to predict whether a fluid will flow in a laminar or turbulent regime.  

The Stokes number, on the other hand, is a dimensionless quantity used to describe the behavior of particles in a fluid. It is calculated by dividing the characteristic time scale of a particle's motion by the characteristic time scale of the fluid flow. A low Stokes number indicates that the particles are strongly affected by the fluid flow, while a high Stokes number indicates a negligibly impact of fluid flow on the dynamics of particles. The Stokes number is defined as \cite{RDS16}
\beq
\label{VII2}
\text{St}=\frac{\rho_p(1-\phi)\sigma\Delta\mathbf{U}}{18\eta_g},
\eeq
where $\rho_p$ is the particle density. Considering $m=(\pi/6)\sigma^3\rho_p$ for $d=3$, the Stokes number reads
\beq
\label{VII2bis}
\text{St}=\frac{(1-\phi)}{\sigma\gamma_{\text{St}}}\Delta\mathbf{U}.
\eeq
If $\Delta\mathbf{U}= v_\text{th}= \sqrt{2T/m}$, Eq.\ \eqref{VII2bis} can be written as
\beq\label{VII5}
\text{St}=  \frac{1}{12}\sqrt{\frac{\pi}{2}}\, \frac{1-\phi}{\phi}\, \frac{R}{\chi}\, \frac{1}{\gamma^*}
\eeq

In this paper we have considered a scenario where the effect of inelasticity in collisions on the dynamics of grains is comparable to that of the interstitial gas, which means that our intruder is neither a Brownian particle suspended in a fluid nor an intruder in a dry granular gas. For this reason, we are interested in granular suspensions where $\nu/\gamma_{\text{St}}\sim 1/\gamma^*\sim 1$. Note that $\nu/\gamma_{\text{St}}$ is an estimate of the number of collisions of the grains during the Langevin relaxation time $1/\gamma_{\text{St}}$.

Let us consider a Brownian spherical particle with a diameter $\sigma = 10$ nm  immersed in air at normal temperature (25 ºC) and pressure (1  atm) where $\eta_g(\text{air})= 1.8\times 10^{-5}$ Pa$\cdot$s. These conditions can be considered as representative of an ordinary  state of the interstitial molecular gas. Using Eqs.\ \eqref{VII1} and \eqref{VII2bis} with $\Delta\mathbf{U}=v_\text{th}$, the Reynolds and Stokes numbers are approximately $3 \times 10^{-3}$ and $0.1$, respectively.  These values are consistent with the approximations we made. However, in this case,   $T_\text{b}^*\sim 10^{-2}$, which is a very small value far from the choice $T_\text{b}^*=1$ used in our graphs and simulations.

According to Eqs.\ \eqref{2.4.0} and \eqref{2.13.1}, since $\mathcal{T}\propto \sigma \eta_g^2/\rho_p^2$, one way to increase the value of  $T_\text{b}^*=T/\mathcal{T}$ is by decreasing the value of $\sigma$.
Another way, of course, is to consider a molecular gas with a lower viscosity $\eta_g$ and a denser grain $\rho_p$. For example, we can choose hydrogen  as the molecular gas ($\eta_g=8.8\times 10^{-6}$ Pa$\cdot$s at normal temperature and pressure) and gold particles as grains ($\rho_p = 1.93\times 10^4$ kg/m$^3$). In this case, for $\sigma=10$ nm, one has $T_\text{b}^*=0.6$, a value already close to $T_\text{b}^*=1$. Table\ \ref{table1} shows Re, St, $\gamma_{\text{St}}/\nu$ and $T_\text{b}^*$ values for other sizes of the gold grain. 
Note that these parameters take values compatible with the approximations and assumptions made along the paper. 
While the Reynolds numbers are of order $10^{-4}$ or less, the Stokes numbers, $T_\text{b}^*$, and $\gamma_\text{St}/\nu$ are close to unity. Moreover, any change in the granular temperature by a reasonable factor does not substantially affect the values of  Re, St and $\gamma_\text{St}/\nu$, as the lower part of table\ \ref{table1} shows.

\begin{table}
\begin{tabular}{|c|c|c|c|c|c|c|c|}
 \hline
  $\sigma$ & $\phi$ & Re   & St &  $\gammaSt/\nu$  & $T_\text{b}^*$ \\
  \hline 
  \hline
1 nm  & 0.1   & $3\times 10^{-4}$ &3.1  & 0.2  & 6 \\
10 nm & 0.1   & $8\times 10^{-5}$ &1.0  & 0.7  & 0.6\\
100 nm & 0.1   & $3\times 10^{-5}$ &0.3  & 2.3  & 0.06 \\
1 nm & 0.2   & $2\times 10^{-4}$ &2.8  & 0.1 & 6 \\
10 nm & 0.2   & $7\times 10^{-5}$ &0.9 & 0.3  & 0.6 \\
100 nm & 0.2   & $2\times 10^{-5}$ &0.3  & 0.9 & 0.06 \\
\hline 
\hline
 1 nm  & 0.1   & $2\times 10^{-4}$ &2.2 &  0.3 & 6 \\
 10 nm & 0.1   & $6\times 10^{-5}$ &0.7 &  1.0  & 0.6\\
 1 nm & 0.2   & $2\times 10^{-4}$ &2.0 &  0.1  & 6 \\
 10 nm & 0.2   & $5\times 10^{-5}$ &0.6 &  0.4  & 0.6 \\
  \hline
\end{tabular}
\caption{
Table of various parameters for a suspension of gold grains  immersed in hydrogen molecular gas at normal temperature,  298.15 K, and pressure $1.01\times 10^5$ Pa. Several diameters and volume fractions of the gold grains are considered. In the upper part of the table, both the bath temperature $T_\text{b}$ and the granular temperature $T$ take a common value (corresponding to the normal temperature 298.15 K). The entries in the lower part of the table are parameter values corresponding to the same $T_\text{b}$-value as in the upper part, but with 
$T = T_\text{b}/2$.}
\label{table1}
\end{table}

The Stokes--Einstein formula \eqref{DStoEin} [or equivalently, $D_\text{SE}^*=T_\text{b}^*$; see Sec.\ \ref{secRWSelf}] provides the self-diffusion coefficient of an isolated grain in a suspension. If the grain is surrounded by other mechanically equivalent grains with concentration $\phi$, an estimate of the self-diffusion coefficient $D$ in the first Sonine approximation is given by Eq.~\eqref{4.22}. Moreover, Figs.\ \ref{fig3}, \ref{fig6}, and \ref{fig7} clearly show that the effect of inelasticity in collisions on the diffusion coefficient $D$ is in general very weak. This means that the functional form of $D$ for elastic and inelastic collisions is almost the same, as long as 
$\alpha$ is not too small.

For elastic collisions and mechanically equivalent particles, the relation between the self-diffusion coefficients $D$ and $D_\text{SE}$ is
\beq
\label{n1}
\frac{D}{D_\text{SE}}=\frac{1}{R+16 \phi\chi \sqrt{T_\text{b}^*/\pi}},
\eeq
where we recall that $R(\phi)$ accounts for the density dependence of the friction coefficient $\gamma$ (c.f., Eq.~\eqref{2.5}). The other term ($16 \phi\chi \sqrt{T_\text{b}^*/\pi}$) of the denominator of Eq.\ \eqref{n1} accounts for the collisions between grains. For $\phi=0.1$ and $T_\text{b}^*=0.06$, we find $16 \phi\chi \sqrt{T_\text{b}^*/\pi}\approx 0.3$, which is small compared to $R(0.1)=2.2$. Therefore, the diffusion coefficient is mainly determined by the interaction of the grain with the interstitial gas. 
The impact of collisions on diffusion increases with  $T_\text{b}^*$. For example, for $\phi=0.1$ and $T_\text{b}^*=6$, we find $16 \phi\chi \sqrt{T_\text{b}^*/\pi}\approx 2.8$, that is of the same order as $R(0.1)=2.2$. The joint effect of both terms leads to a change of the ratio $D/D_\text{SE}$ by a factor of five with respect to the Stokes--Einstein value. 
Thus, we find that in general the influence of grain-grain collisions on the diffusion coefficient is non-negligible for most of the cases considered in table \ref{table1}.
It must be noted that this conclusion is robust against the particular choice for $R$ or $\chi$. For example, had we used the extreme values $R=1$ and $\chi=1$ (i.e., the values corresponding to the dilute limit), the ratio $D/D_\text{SE}$ would still have changed by a factor of three instead of five

In summary,  grain-grain
collisions modify in general the diffusion coefficient by a large percentage, highlighting the importance of incorporating collision effects into models of granular suspensions.  


\section{Summary and outlook}
\label{sec7}

Let us recap the main results and the methodology employed throughout this paper. We have used the Chapman--Enskog method to solve the Enskog--Lorentz kinetic equation  up to the first order in the density gradient. From this solution we have obtained the integral equation obeyed by the diffusion coefficient $D$ of an intruder immersed in a granular suspension of smooth inelastic hard spheres (grains). As in the case of elastic collisions \cite{CC70}, the above integral equation can be solved by expanding the distribution function in a series of Sonine polynomials. Here, we have truncated the series by considering the two first relevant Sonine polynomials. This yields the so-called first and second Sonine approximations to the diffusion coefficient. These sort of solutions have allowed us to find a rich phenomenology for the coefficient $D$, which is in fact a measure of the MSD up to a given time.  

To test the reliability of the Sonine approximations, we have numerically solved the Enskog--Lorentz equation by means of the DSMC method, conveniently adapted to account for inelastic collisions. Although the first-Sonine approximation to $D$ yields in general a good agreement with simulations, we have shown that it is outperformed by the second-Sonine approximation, especially when the intruder is much lighter than the particles of the granular gas. This conclusion agrees with previous findings reported for dry granular mixtures \cite{GM04,GV09,GV12}. However, the influence of inelasticity on mass transport here is weaker than in the absence of the gas phase.

 Although our theoretical results have been derived for arbitrary values of the coefficients of normal restitution $\al$ (for grain-grain collisions) and $\al_0$ (for intruder-grain collisions), we have assumed a common coefficient of restitution ($\al=\al_0$) for the sake of illustration. In this case, we find a non-monotonic behavior of the MSD as a function of $\al$ which is enhanced for sufficiently high density of the granular gas and/or temperature of the interstitial fluid. A similar behavior had already been found in a suspension model with $\gamma=\gamma_0\equiv\text{const}$ \cite{SVCP10} and in the case of a dry granular gas \cite{ABG22}. As in this latter case (see Ref.\ \cite{ABG22}), this effect can be intuitively understood with the help a random walk model allowing one to write the intruder's MSD as the number of collisions with the grains (jumps in the random walk model) multiplied by the square of an EMFP [cf.\ Eq.\ \eqref{rw-msd}] This EMFP accounts for the \emph{positive} correlations between the precollisional and the post-collisional trajectories of the intruder [cf.\ \eqref{sq-EMFP}], and is therefore larger than the actual MFP. The EMFP decreases with increasing $\al$, reflecting a reduction in the persistence of the intruder's motion that is detrimental to the MSD. In contrast, the collision frequency (and thus the number of steps up to a given time) increases strongly with $\al$
 in the quasielastic regime, and the resulting competition with the EMFP leads to the aforementioned non-monotonic behavior.

 For fixed $\al$,  the intricate dependence of the MSD on intruder's mass and diameter is determined by the dependence of the collision frequency and the EMFP on those quantities. The collision frequency grows with the intruder's diameter, but is found to decrease when the intruder becomes heavier. In contrast, the EMFP is found to increase with the mass of the intruder and to decrease when its diameter grows.

Finally, in view of our results in Sec. \ref{secAppli}, one of the main conclusions of this paper would be that, in general, collision effects may have a crucial influence on the behavior of real suspensions, and therefore deserve to be included in the models as a general working principle.

As stated in Secs.\ \ref{sec2} and \ref{sec3}, the theoretical results reported in this paper have been obtained from a coarse-grained approach where the effect of the interstitial fluid on grains has been accounted for via a fluid-solid force. It would be interesting to revisit the tracer diffusion problem by considering a collisional model that explicitly takes into account not only the collisions between grains and particles of the surrounding molecular gas, but also those between the intruders and molecular gas particles. Such a system will thus involve three phases. This sort of collisional suspension model has been recently used \cite{GG22} for the study of gas-solid flows involving two phases; the results derived from this collisional model have been shown to reduce to those derived from the Langevin-like approach \cite{GGG19a} when the grains are much heavier than the particles of the background gas. We expect that a similar conclusion can be achieved in the tracer diffusion problem analyzed in the present paper. Our model could also be extended in other directions, e.g., by introducing additional restitution coefficients to account for boundary and rugosity effects (rough spheres), or by considering binary granular mixtures with arbitrary concentration. Last but not least, it would also be desirable to perform molecular dynamics simulations to assess the reliability of the Enskog kinetic equation. We plan to address these problems in the near future.

\acknowledgments

We acknowledge financial support from Grant PID2020-112936GB-I00 funded by MCIN/AEI/10.13039/501100011033, and 
from Grant IB20079 funded by Junta de Extremadura (Spain) and by ERDF A way of making Europe. The research of R.G.G. also has been supported by Plan
Propio de Iniciación a la Investigación, Desarrollo Tecnológico e Innovación de la
Universidad de Extremadura (ACCIÓN III).

\appendix
\section{First and second Sonine approximations to the diffusion coefficient}
\label{appA}

In this appendix we give some technical details on the determination of the Sonine coefficients $a_1$ and $a_2$. Substitution of 
Eq.\ \eqref{4.12} into the integral equation \eqref{4.8} yields
\begin{widetext}
\beq
\label{a1}
\gamma_0\frac{\partial}{\partial\mathbf{v}}\cdot\mathbf{v} \Big(a_1 f_{0\text{M}}\mathbf{v}+a_2 f_{0\text{M}}\mathbf{S}_0\Big)
+\frac{\gamma_0 T_{\text{b}}}{m_0}\frac{\partial^2}{\partial v^2}\Big(a_1 f_{0\text{M}}\mathbf{v}+a_2 f_{0\text{M}}\mathbf{S}_0\Big)+a_1J_0[f_{0\text{M}}\mathbf{v},f]+a_2J_0[f_{0\text{M}}\mathbf{S}_0,f]=-\frac{f_0^{(0)}}{n_0} \mathbf{v}.
\eeq
\end{widetext}
Next, we multiply Eq.\ \eqref{a1} by $\mathbf{v}$ and integrate over the velocity. The result is
\beq
\label{a2}
\left(\gamma_0+\nu_a\right)D+\frac{n_0 T_0^2}{m_0} \nu_b a_2=\frac{T_0}{m_0},
\eeq
where use has been made of the identity $a_1=(m_0 D/n_0 T_0)$ and have introduced the quantities
\beq
\label{a3}
\nu_a=-\frac{m_0}{d n_0 T_0}\int d\mathbf{v}\; \mathbf{v}\cdot J_0[f_{0\text{M}}\mathbf{v},f],
\eeq
\beq
\label{a4}
\nu_b=-\frac{m_0}{d n_0 T_0^2}\int d\mathbf{v}\; \mathbf{v}\cdot J_0[f_{0\text{M}}\mathbf{S}_0,f].
\eeq
If only the first Sonine corrections is retained ($a_2=0$), the solution to Eq.\ \eqref{a2} is
\beq
\label{a5}
D[1]=\frac{T_0/m_0}{\gamma_0+\nu_a}.
\eeq
Equation \eqref{a5} leads to Eq.\ \eqref{4.19} when the definition $D^*[1]=D[1]/(\gamma_\text{St}\sigma^2)$ is considered.

To close the problem, one has to multiply Eq.\ \eqref{a2} by $\mathbf{S}_0(\mathbf{v})$ and integrate over $\mathbf{v}$. After some algebra, one is left with
\beq
\label{a6}
\frac{m_0}{n_0 T_0^2}\Big[2\gamma_0\Big(1-\frac{T_\text{b}}{T_0}\Big)+\nu_c\Big]D+\left(3\gamma_0+\nu_d\right)a_2=0,
\eeq
where
\beq
\label{a7}
\nu_c=-\frac{2}{d(d+2)}\frac{m_0}{n_0 T_0^2}\int d\mathbf{v}\; \mathbf{S}_0\cdot J_0[f_{0\text{M}}\mathbf{v},f],
\eeq
\beq
\label{a8}
\nu_d=-\frac{2}{d(d+2)}\frac{m_0}{n_0 T_0^3}\int d\mathbf{v}\; \mathbf{S}_0\cdot J_0[f_{0\text{M}}\mathbf{S}_0,f].
\eeq

In reduced units and using matrix notation, Eqs.\ \eqref{a2} and \eqref{a6} can be rewritten as
\beq
\label{a9}
\left(
\begin{array}{cc}
(\gamma_0^*+\nu_a^*)\xi^*&\tau_0^2 \nu_b^*\\
\xi^*\frac{\nu_c^*+2\gamma_0^*\left(1-\frac{T_\text{b}^*}{T_0^*}\right)}{\tau_0^2}&3\gamma_0^*+\nu_d^*
\end{array}
\right)
\left(
\begin{array}{c}
D^*\\
a_2^*
\end{array}
\right)
=
\left(
\begin{array}{c}
\tau_0\\
0
\end{array}
\right).
\eeq
Here, $\tau_0=T_0/T$, $D^*=D/(\gamma_{\text{St}}\sigma^2)$, $\xi^*=m_0 R/(m T^* \gamma^*)$, $a_2^*=n_0 T \nu a_2$, $\nu_a^*=\nu_a/\nu$, $\nu_b^*=\nu_b/\nu$, $\nu_c^*=\nu_c/\nu$, and $\nu_d^*=\nu_d/\nu$. The reduced friction coefficients $\gamma^*$ and $\gamma_0^*$ are defined by Eq.\ \eqref{4.20}, while the effective collision frequency $\nu$ is defined by Eq.\ \eqref{2.11}. The solution to Eq.\ \eqref{a9} gives the expression of the second Sonine approximation $D^*[2]$ to $D^*$, which reads as follows
\beq
\label{a11}
D^*[2]=\frac{\xi^{*-1}\tau_0(\nu_d^*+3\gamma_0^*)}{(\nu_a^*+\gamma_0^*)(\nu_d^*+3\gamma_0^*)-\nu_b^*\Big[\nu_c^*+2\gamma_0^*
\Big(1-\frac{T_\text{b}^*}{T_0^*}\Big)\Big]}.
\eeq
The expression \eqref{a11} yields directly Eq.\ \eqref{4.17} for the second-Sonine approximation to $D^*$.

To obtain the explicit dependence of $D^*[2]$ and $D^*[1]$ on the parameter space of the system, one still needs to determine the quantities $\nu_a^*$, $\nu_b^*$, $\nu_c^*$, and $\nu_d^*$. These quantities have been evaluated in previous works \cite{GM07,GHD07,GV09} when the distribution $f$ is approximated by the Maxwellian distribution \eqref{2.9.1}. We reproduce the explicit expressions below:
\begin{widetext}
\begin{equation}
\label{a12}
\nu_{a}^*=\frac{\sqrt{2}}{d}\left(\frac{\overline{\sigma}}{\sigma}
\right)^{d-1}\frac{\chi_0}{\chi}\mu (1+\alpha_0) \left(\frac{1+\beta}{\beta}\right)^{1/2},
\end{equation}
\begin{equation}
\label{a13}
\nu_{b}^*=\frac{1}
{\sqrt{2}d}\left(\frac{\overline{\sigma}}{\sigma}\right)^{d-1}\frac{\chi_0}{\chi}\mu(1+\alpha_0)[\beta(1+\beta)]^{-1/2},
\end{equation}
\begin{equation}
\label{a14}
\nu_{c}^*=\frac{\sqrt{2}}
{d(d+2)}\left(\frac{\overline{\sigma}}{\sigma}\right)^{d-1}
\frac{\chi_0}{\chi}\mu(1+\alpha_0)\left(\frac{\beta}{1+\beta}\right)^{1/2}A_c,
\end{equation}
\begin{equation}
\label{a15}
\nu_{d}^*=\frac{1}
{\sqrt{2}d(d+2)}\left(\frac{\overline{\sigma}}{\sigma}\right)^{d-1}
\frac{\chi_0}{\chi}\mu(1+\alpha_0)\left(\frac{\beta}{1+\beta}\right)^{3/2}
\left[A_d-(d+2)\frac{1+\beta}{\beta} A_c\right],
\end{equation}
where
\begin{eqnarray}
\label{a16}
A_c&=& (d+2)(1+2\lambda)+\mu(1+\beta)\Big\{(d+2)(1-\alpha_0)
-[(11+d)\alpha_0-5d-7]\lambda\beta^{-1}\Big\}+3(d+3)\lambda^2\beta^{-1}\nonumber\\
& &+2\mu^2\left(2\alpha_0^{2}-\frac{d+3}{2}\alpha
_{12}+d+1\right)\beta^{-1}(1+\beta)^2- (d+2)\beta^{-1}(1+\beta),
\end{eqnarray}
\begin{eqnarray}
\label{a17}
A_d&=&2\mu^2\left(\frac{1+\beta}{\beta}\right)^{2}
\left(2\alpha_0^{2}-\frac{d+3}{2}\alpha_0+d+1\right)
\big[d+5+(d+2)\beta\big]-\mu(1+\beta) \Big\{\lambda\beta^{-2}[(d+5)+(d+2)\beta]
\nonumber\\
& & \times
[(11+d)\alpha_0
-5d-7]-\beta^{-1}[20+d(15-7\alpha_0)+d^2(1-\alpha_0)-28\alpha_0] -(d+2)^2(1-\alpha_0)\Big\}
\nonumber\\
& & +3(d+3)\lambda^2\beta^{-2}[d+5+(d+2)\beta]+ 2\lambda\beta^{-1}[24+11d+d^2+(d+2)^2\beta]
\nonumber\\
& & +(d+2)\beta^{-1} [d+3+(d+8)\beta]-(d+2)(1+\beta)\beta^{-2}
[d+3+(d+2)\beta].\nonumber\\
\end{eqnarray}
Here, $\lambda=(\mu_0/T_0)\left(T_0-T\right)$.
\end{widetext}


%

\end{document}